\documentclass[aps,prb,twocolumn,amsmath,amssymb,superscriptaddress,longbibliography]{revtex4-2}

\usepackage{bm}

\usepackage{graphicx}
\usepackage{caption}
\usepackage{subcaption}

\usepackage{lmodern}
\usepackage{xcolor}
\usepackage{changes}
\usepackage[normalem]{ulem}
\usepackage{float}
\usepackage{braket}
\usepackage{dsfont}
\usepackage{mathtools}
\usepackage{tabularx}
    \newcolumntype{L}{>{\raggedright\arraybackslash}X}

\usepackage{xcolor}
\usepackage[bookmarksnumbered]{hyperref}
\definecolor{cblue}{RGB}{55,126,184}
\hypersetup{colorlinks=true,linkcolor=cblue,citecolor=cblue,urlcolor=cblue}

\parskip=\baselineskip
\def\be{\begin{equation}}
\def\ee{\end{equation}}

\newcommand{\beq}{\begin{equation}}
\newcommand{\eeq}{\end{equation}}

\DeclarePairedDelimiterX\bbrakket[2]{\langle\langle}{\rangle\rangle}{#1 \delimsize\Vert #2}

\begin{document}

\title{
Hollow Lattice Tensor Gauge Theories with Bosonic Matter
}

\author{Jos\'{e} M. Cruz}
\affiliation{CeFEMA-LaPMET, Departamento de F\'{ı}sica, Instituto Superior T\'{e}cnico, Universidade de Lisboa, Av. Rovisco Pais, 1049-001 Lisboa, Portugal}
\affiliation{Department of Physics, Gakushuin University, Mejiro, Tokyo, 171-8588, Japan}
\affiliation{Department of Physics, University College Dublin, Dublin, D04V1W8, Ireland}
\email{jose.m.cruz@ucdconnect.ie}
\author{Masafumi Udagawa}
\affiliation{Department of Physics, Gakushuin University, Mejiro, Tokyo, 171-8588, Japan}
\author{Pedro Bicudo}
\affiliation{CeFEMA-LaPMET, Departamento de F\'{ı}sica, Instituto Superior T\'{e}cnico, Universidade de Lisboa, Av. Rovisco Pais, 1049-001 Lisboa, Portugal}
\author{Pedro Ribeiro}
\affiliation{CeFEMA-LaPMET, Departamento de F\'{ı}sica, Instituto Superior T\'{e}cnico, Universidade de Lisboa, Av. Rovisco Pais, 1049-001 Lisboa, Portugal}
\affiliation{Beijing Computational Science Research Center, Beijing 100193, China}
\author{Paul~A.~McClarty}
\affiliation{Laboratoire L\'{e}on Brillouin, CEA, CNRS,  Universit\'{e} Paris-Saclay, CEA-Saclay, 91191 Gif-sur-Yvette, France}
\affiliation{ Max Planck Institute for the Physics of Complex Systems, N\"{o}thnitzer Str. 38, 01187 Dresden, Germany}
\email{paul.mc-clarty@cnrs.fr}

\begin{abstract}
Higher rank gauge theories are generalizations of electromagnetism where, in addition to overall charge conservation, there is also conservation of higher rank multipoles such as the total dipole moment. In this work we study a four dimensional lattice tensor gauge theory coupled to bosonic matter which has second rank tensor electric and magnetic fields and charge conservation on individual planes. 
Starting from the Hamiltonian, we derive the lattice action for the gauge fields coupled to $q=1,2$ charged scalars. We use the action  formulation to carry out Monte Carlo simulations to map the phase diagram as a function of the 
gauge ($\beta$) and matter ($\kappa$) couplings.  
We compute the nature of correlators at strong and weak coupling in the pure gauge theory and compare the results to numerical simulations. Simulations show that the naive weak coupling regime (small $\kappa$, large $\beta$) does not survive in the thermodynamic limit. Instead, the strong coupling confined phase, spans the whole phase diagram. It is a proliferation of instantons that destroys the weak coupling phase and we show, via a duality transformation, that the expected strong confinement is present in the analog of Wilson line correlators. 
For finite matter coupling at $q=1$ we find a single thermodynamic phase albeit with a first order phase transition terminating in a critical endpoint.
For $q=2$ it is known that the the X-cube model with $\mathbb{Z}_2$ fractonic topological order is recovered deep in the Higgs regime. The simulations indeed reveal a distinct Higgs phase in this case.
\end{abstract}

\maketitle

\tableofcontents

\vspace{-0.7cm}
\section{Introduction} 

\vspace{-0.5cm}

Condensed matter systems have long furnished physicists with a tremendous variety of distinct states of matter and it has been the work of many decades to find organizing principles general enough to encompass them. Following the work of Wilson \cite{Wilson1974} and Polyakov \cite{Polyakov1977} on strongly coupled gauge theories on the lattice, it became necessary to generalize older ideas of symmetry breaking and local order parameters that were the staple of traditional solid state physics. Phases characterized by nonlocal order parameters and higher form symmetries arise naturally among certain classes of condensed matter systems such as quantum spin liquids where the low energy physics is captured by some kind of gauge principle. 

Among such unusual states of matter are those with exotic-seeming conservation laws such as conservation of dipole moments \cite{Pretko_2017,Pretko_2017_2}. Consideration of models founded on such conservation laws came in the wake of the discovery of fracton phases  \cite{haahcode2011,bravyi2013,vijay2015,vijayhaahfu2016} where quasiparticles arise with restricted mobility. For more details on fractons, together with references to the quite extensive literature, we refer the reader to reviews \cite{Nandkishore_2019,Pretko_2020,gromov2024,you2024quantumliquidsemergenthigherrank} on the subject. Following the discovery of such states of matter it was natural to look for gauge theories that are to these higher rank conservation laws as ordinary U(1) gauge theory is to charge conservation. 

The resulting counterparts to ordinary Maxwell electrodynamics, higher rank gauge theories, have symmetric tensorial gauge fields and a generalized Abelian gauge invariance \cite{Pretko_2017,Pretko_2017_2,gu2006,xucenke2006,wu2xu008,Rasmussen_2016,slagle2019,du2021,hermele2021,Seiberg_2020,Gorantla_2021,Seiberg2021,Seiberg2022,Seiberg2023}. In addition to their roots in fracton phases they have connections to certain classical constrained systems with glassy dynamics and elastic media \cite{PhysRevB.95.155133,PhysRevLett.120.195301,gromov2024}. Their weak coupling continuum description also carries surprises by containing short wavelength low energy modes thus breaking the usual clean separation between UV and IR physics \cite{Seiberg_2020}. 

Our interest, in this paper, is to provide a global view of the phase diagram of lattice-regularized higher rank gauge theories in cases where the naive continuum limit breaks down \cite{xucenke2008}. To this end we provide a combined numerical and analytical study of the hollow rank-two theory in four dimensions \cite{wu2xu008,Ma_2018,Bulmash_2018}, otherwise known as A tensor gauge theory \cite{Seiberg_2020} coupled to scalar matter. We present results of Monte Carlo simulations as a function of both gauge $\beta$ and matter $\kappa$ couplings for both charge one $q=1$ and charge two $q=2$ Higgs fields. The latter case is of interest as a limit of the theory has fracton topological order \cite{Ma_2018,Bulmash_2018}. 

This work parallels early work on lattice Abelian Higgs models \cite{fradkinshenker1979,PhysRevD.25.531}  where the strong coupling pure gauge theory was shown to exhibit a confined phase. In $3+1$-dimensions, in the pure U(1) gauge theory, there is a phase transition to a deconfined phase where Maxwell electromagnetism describes the low energy physics. The confined and deconfined phases of the pure gauge theory can be characterized by, respectively, an area and perimeter law for Wilson line operators. Coupling to scalar matter replaces the area law by a perimeter law because the Wilson line is screened by the charged matter. But the deconfined phase has power law correlations while the confined phase that smoothly connects to a Higgs regime is gapped. 

For the pure gauge theory, we show that the finite size simulation results are compatible with weak and strong coupling expansions. However, finite size scaling reveals that features from the naive continuum limit at weak coupling do not survive the thermodynamic limit. As we know from earlier work \cite{xucenke2008}, proliferation of instantons destroy the analog of the deconfined phase. We revisit this issue from a modern perspective by passing to dual variables. As we show in detail, the instantons that disorder the gauge fields lead to behavior that qualitatively matches the strong coupling behavior indicating a single phase for all $\beta$. This is the higher rank analog of the Polyakov confinement mechanism present in pure U(1) gauge theory in $2+1$ dimensions and in {\it two-form } gauge theory in $3+1$ dimensions \cite{polyakovGaugeFieldsStrings2017}. Pinch points, that would be expected in certain electric field correlation functions from the  continuum tensor gauge field theory \cite{prem2018}, are absent in the quantum phase diagram. 

From these considerations combined with simulation results we conclude that the Higgsed A tensor gauge theory with charge $q=1$ has a single phase passing from Higgs to confined regimes but with a line of first order phase transitions terminating at a critical endpoint reminiscent of a liquid-gas phase diagram. 

In contrast, in the $q=2$ theory, there are two phases separated by a first order line. One of these phases $-$ that we may call the Higgs phase $-$ is continuously connected to the X-cube model \cite{Ma_2018,Bulmash_2018} and therefore has fractonic topological order.

\vspace{-0.5cm}

\section{Hollow Rank-2 Gauge Theory with Bosonic Matter}

\vspace{-0.5cm}

\subsection{Hamiltonian and Lattice Action}

\vspace{-0.5cm}

A natural starting point to formulate the lattice gauge theory is first to introduce global symmetries associated with currents $J_0$ and $J_{ij}$, that is symmetric in its indices $i,j \in \{x,y,z\}$ and $i\neq j$, satisfying continuity equations
\be
\partial_0 J_0 = \sum_{i<j} \partial_i\partial_j J_{ij} 
\ee
implying conserved charges on planes. We may gauge these global symmetries by introducing gauge fields $A_0 \rightarrow A_0 + \partial_0 \Omega$ and  $A_{ij} \rightarrow A_{ij} + \partial_i \partial_j \Omega$. The minimal coupling in the action $A_0 J_0 + A_{ij} J_{ij}$ gauges these global symmetries.

We then pass to the lattice by placing $U(1)$ rotor variables on plaquettes of a three dimensional cubic lattice indexed by vertices $\vec{r}$. Spatial plaquettes or faces are indexed by $\vec{r}$ and $ij=xy, xz, yz$. On each spatial plaquette there is a phase variable $A_{ij}(\vec{r})$ and conjugate number operator denoted $E_{ij}(\vec{r})$ obeying
\be
\left[ A_{ij}(\vec{r}),  E_{ij}(\vec{r}) \right] = i \delta_{\vec{r},\vec{r}'} \left(  \delta_{ik}\delta_{jl} + \delta_{il}\delta_{jk}  \right).
\ee
The lattice Gauss law is
\be
\sum_{i< j} \Delta_i \Delta_j E_{ij} = \rho.
\ee
where $\rho$ is the scalar matter density defined on vertices and about which we say more below. 

We further introduce operators $U_{ij}(\vec{r}) \equiv \exp(i a^2 A_{ij}(\vec{r}) )$. Gauge transformations act on these as
\begin{align}
U_{xy}(\vec{r}) \rightarrow& U_{xy}(\vec{r})\eta(\vec{r})\eta^{-1}(\vec{r}+\vec{e_x}) \\
&\eta(\vec{r}+\vec{e_x}+\vec{e_y})\eta^{-1}(\vec{r}+\vec{e_y}) \nonumber
\label{eq:AijGaugeTransformation}
\end{align}
where $\eta(\vec{r})$ is a phase shift acting on a vertex. This is the lattice version of the transformation of continuum $A_{xy}$ given above and which has a straightforward generalization to the other plaquette operators. 

\begin{figure}[H]
    \centering
    \includegraphics[width=\columnwidth]{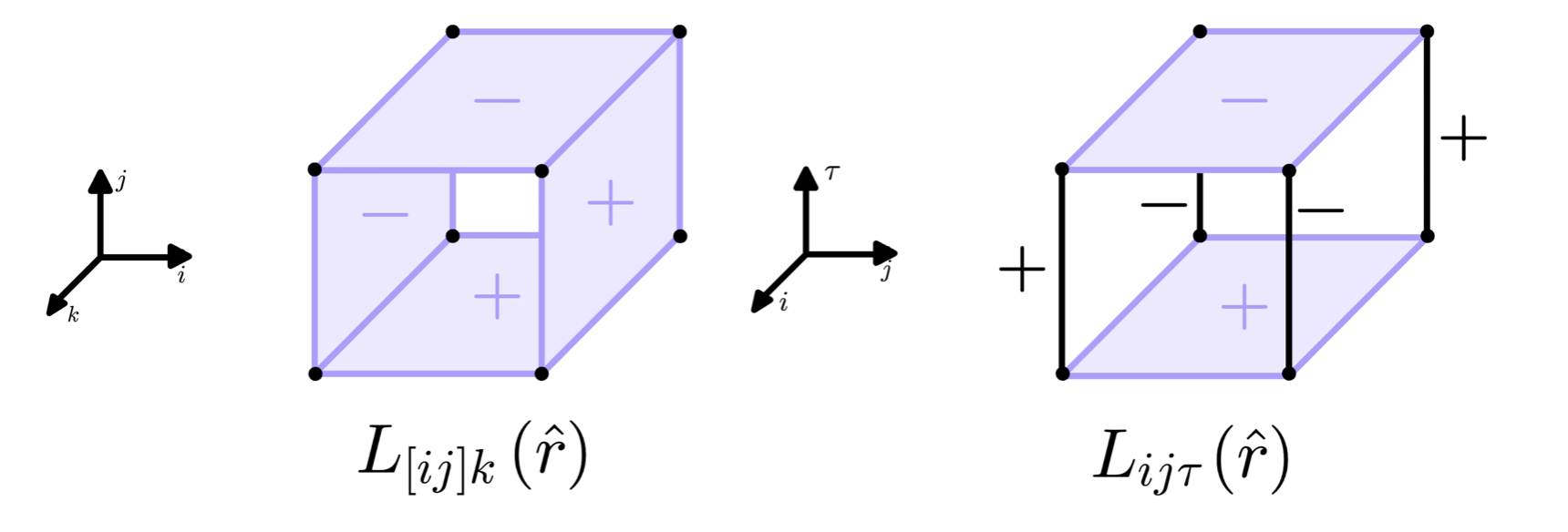}
    \caption{Representation of operators $L_{[ij]k}$ and $L_{ij\tau}$ introduced in the main text. }
    \label{fig:gauge_inv}
\end{figure}

From these fields we may build gauge invariant local field strengths $L_{[ij]k}(\vec{r})$ with $i\neq j\neq k$:
\be
L_{[ij]k}=\exp \left(i( \Delta_i A_{jk}-\Delta_j A_{ik})\right)
\ee
supported on elementary tubes on the lattice (as illustrated in the right panel of Fig.~\ref{fig:gauge_inv}).      

\vspace{-0.5cm}

The Hamiltonian for the pure  lattice gauge theory is 
\be
H_{\rm Gauge} = \sum_{\vec{r}, i<j} \frac{1}{g_e^2} E_{ij}^2+\frac{1}{g_m^2}( L_{[ij]k}(\vec{r})+L_{[ij]k}^\dag(\vec{r})).
\ee

In the literature, this theory goes by several names: the $A$ tensor gauge theory \cite{Seiberg_2020}, the rank-$2$ hollow $U(1)$ gauge theory \cite{Ma_2018}, and the $(0,1)-$scalar rank-$2$ gauge theory \cite{Bulmash_2018}.

In addition we introduce scalar matter fields $\theta(\vec{r})$ coupled to the gauge fields together with conjugate fields $L(\vec{r})$:
\be
[\theta(\vec{r}), L(\vec{r}')] = i \delta_{\vec{r},\vec{r}'}.
\ee
We choose the Higgs fields to have integer charge $q$ such that the Hamiltonian is
\be
H_{\rm Higgs} = \sum_{\vec{r}} \frac{L(\vec{r})^2}{2M} - V \sum_{i<j} \cos(\Delta_i \Delta_j \theta (\vec{r}) + q A_{ij}(\vec{r}))
\label{eq:higgs}
\ee
and the Gauss law is, more precisely, 
\be
\sum_{i< j} \Delta_i \Delta_j E_{ij} =  q L(\vec{r}). 
\ee
The matter-gauge coupling term in the Hamiltonian is gauge invariant under simultaneous local shifts of the phase of $\theta(\vec{r})$ and on the corners of the $A_{ij}(\vec{r})$ plaquette operator. 
Evidently, application of the second term in Eq.~\ref{eq:higgs} to the vacuum results in the creation of a plaquette excitation and alternating charges $\pm q$ around the plaquette. Successive application of such operators on neighbouring plaquettes leads to hopping of pairs of charges, forming a dipole, accompanied by a flux string.

The overall Hamiltonian is $H=H_{\rm Gauge} + H_{\rm Higgs}$. To gain insight into this model and for the purposes of investigating the strongly interacting regime numerically it will be convenient to write down a lattice action. 
\begin{align}
    S_{\rm Gauge} = -\beta\sum_{\tau=1}^N  \sum_{\vec{r}}& \cos( \Delta_i A_{jk}(\vec{r})-\Delta_j A_{ki}(\vec{r}) )  \label{eq:gaugelatticeaction} \\
    &+ \cos (\Delta_i \Delta_j A_0^\tau(\vec{r}) -\Delta_0 A_{ij}^\tau (\vec{r})) \nonumber
\end{align}
where the fields are defined on a hypercubic lattice indexed by $\vec{r}=(\hat{\tau},\hat{x},\hat{y},\hat{z})$. The gauge invariant objects in this action are illustrated in Fig.~\ref{fig:gauge_inv}. In addition to the gauge invariant {\it space cubes} $L_{[ij]k}$ (Fig.~\ref{fig:gauge_inv}(left))defined on a spatial cube, there are elementary {\it time cubes} $L_{ij\tau}$ (Fig.~\ref{fig:gauge_inv}(right)) with pairs of spatial plaquettes connected by four temporal links and spatial cubes. 

The Higgs-gauge part of the action is

\vspace{-0.5cm}

\begin{align}
S_{\rm Higgs} = -\kappa\sum_{\tau=1}^N  \sum_{\vec{r}}&  \cos(qA_0^\tau(\vec{r})+\Delta_0 \theta^\tau (\vec{r}))+ \nonumber \\
&\sum_{i<j} \cos(qA_{ij}^\tau(\vec{r})+\Delta_i \Delta_j \theta^\tau(\vec{r}))
\label{eq:higgslatticeaction}
\end{align}

\begin{figure}[H]
    \centering
    \includegraphics[width=0.8\columnwidth]{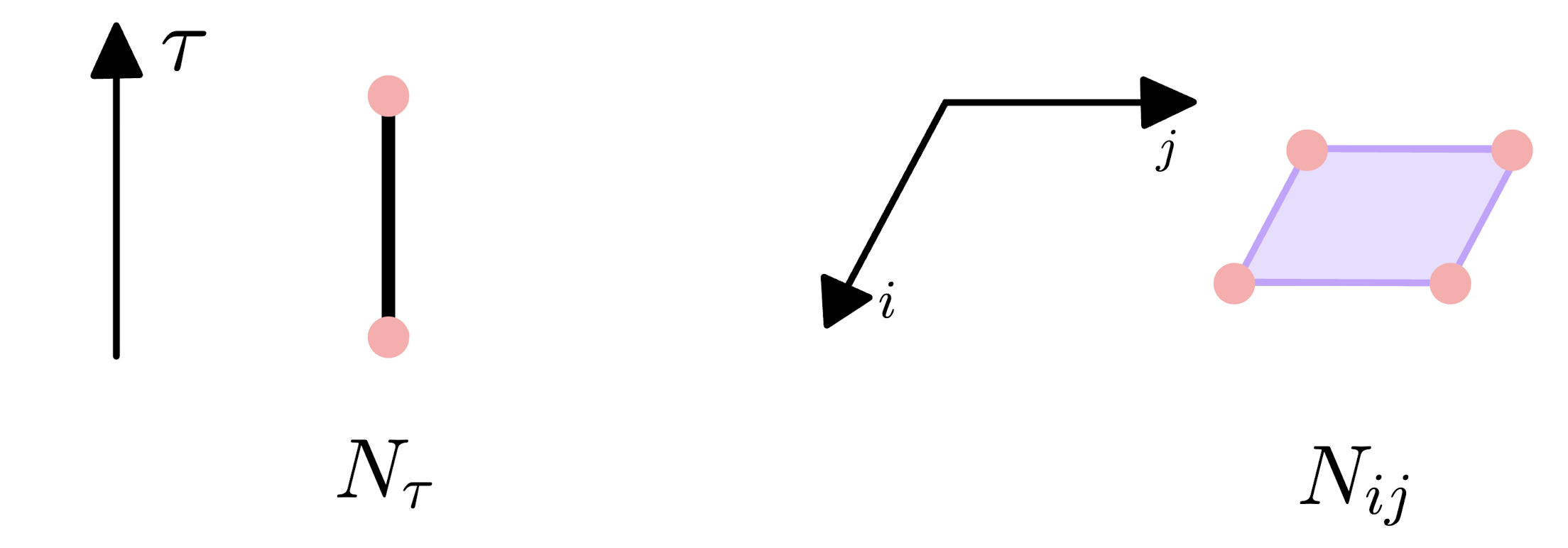}
    \caption{Representation of gauge invariant combinations of gauge and matter fields. On the left we have the time link with matter fields at the ends $N_\tau$. On the right we have the space plaquette with matter fields on the corners $N_{ij}$.}
    \label{fig:gaugeinvmatter}
\end{figure}

The gauge invariant objects in the gauge-matter action are illustrated in Fig.~\ref{fig:gaugeinvmatter}. These are defined on temporal links (left) and on spatial plaquettes (right) with matter fields appearing at the vertices. Under gauge transformations, the action is left invariant by imposing $\theta \rightarrow \theta-q\Omega$. 

\vspace{-0.1cm}

\vspace{-1cm}

A derivation of the action $S=S_{\rm Gauge}+ S_{\rm Higgs}$ from the Hamiltonian is given in appendix~\ref{sec:app_action}.

Somewhat similarly to the generalization of $1$-form to higher form gauge theories on the lattice, the rank-two theory considered here can be generalized to higher rank. For example, instead of plaquette (2-surface) terms $A_{ij}(\vec{r})$ in the rank-two theory, the rank-three theory has variables defined on 3-surfaces.  

\vspace{-0.6cm}

 \subsection{Remarks on the Monte Carlo method}
 \label{sec:montecarlo}

 \vspace{-0.5cm}
 
\begin{figure}[H]
\begin{center}
  \includegraphics[width=0.9\columnwidth]{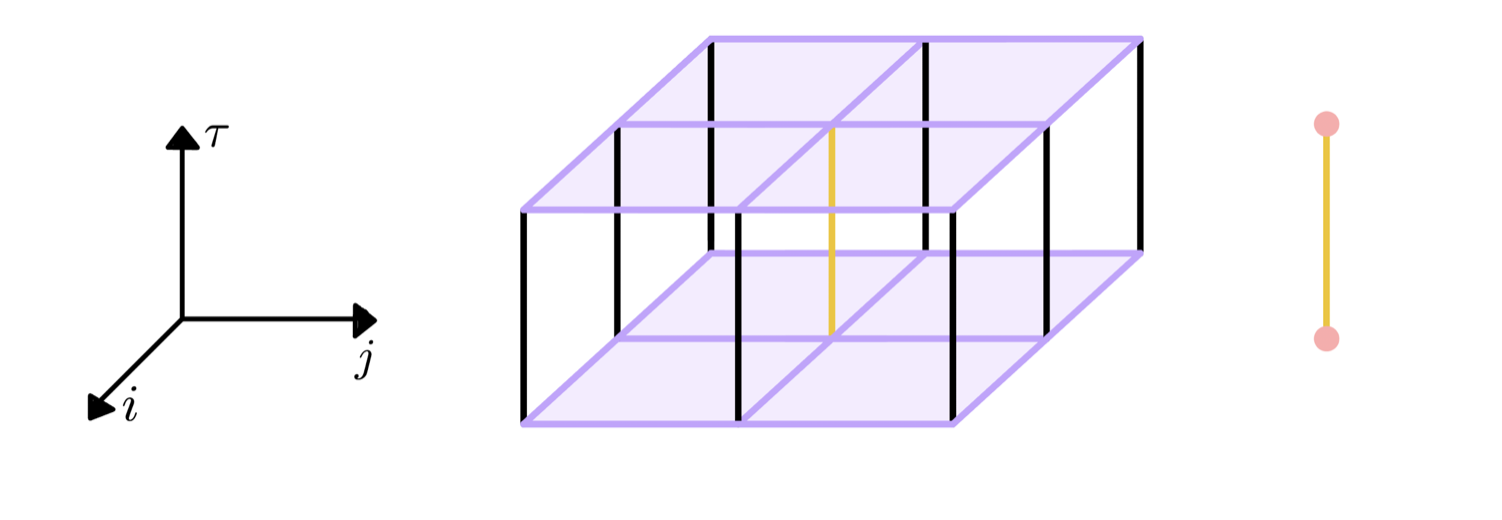}
  \caption{\label{fig:tstaple} Time Staples - staples associated with the temporal link variables. One such variable - the one to be updated - is shown in yellow. Each time link is shared between 12 time cubes $L_{ij\tau}$ (4 for each $ij$ plane) and by one matter $N_\tau$ object.}
  \end{center}
\end{figure}

\begin{figure}[H]
\begin{center}
  \includegraphics[width=1\columnwidth]{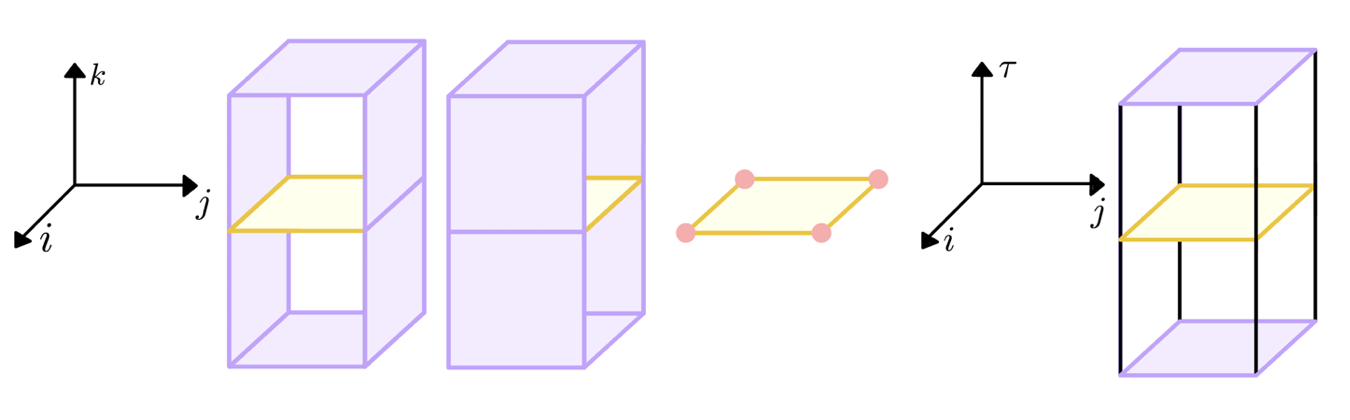}
  \caption{\label{fig:xyzstaple} Space Staples - staples associated with the spatial plaquettes gauge variables. One such variable - the one to be updated - is shown in yellow. Each spatial plaquette is shared between 4 space cubes $L_{[ij]k}$, one matter $N_{ij}$ object and by 2 time cubes $L_{ij\tau}$. }
  \end{center}
\end{figure}

\begin{figure}[H]
\begin{center}
  \includegraphics[width=1\columnwidth]{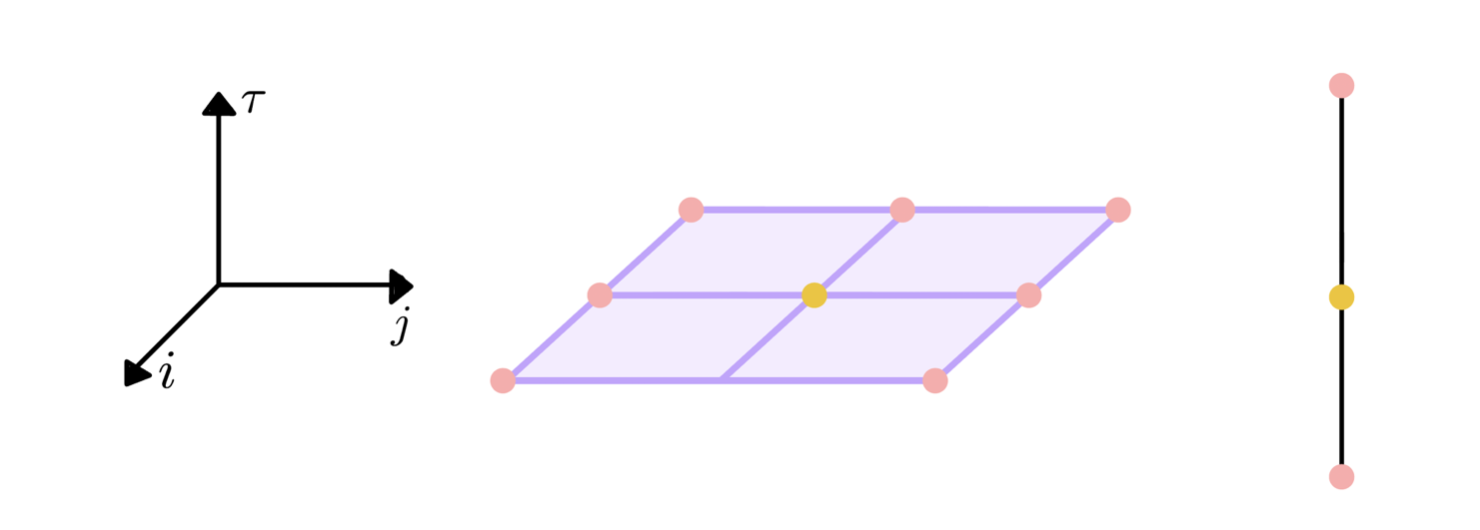}
  \caption{\label{fig:mstaple} Matter Staples - staples associated with the matter fields variables. One such variable - the one to be updated - is shown in yellow. Each matter field is shared between 12 $N_{ij}$ objects (4 for each $ij$ plane) and by 2 $N_\tau$ objects. }
  \end{center}
\end{figure}

 In order to determine the nature of the phases in the theory defined in the previous section, we performed direct Monte Carlo simulations on the lattice action (Eqs.~\ref{eq:gaugelatticeaction} and~\ref{eq:higgslatticeaction}). The Monte Carlo simulations were performed using the Metropolis-Hastings algorithm on a hypercubic $N^4$ lattice with periodic boundary conditions. Efficient updating of the configuration was achieved by using OpenMP and a division of the variables to be updated into disjoint sets labelled by a parity. For temporal links this is a 2-parity, while for spatial plaquettes a 4-parity was used. For each such variable, the lattice is divided into staples $-$ the collection of variables on which the update of a single plaquette, link or vertex variable depends. We show the staples for the update of temporal links (in Fig.~\ref{fig:tstaple}), spatial plaquettes (Fig.~\ref{fig:xyzstaple}) and matter vertices (Fig.~\ref{fig:mstaple}). 

\vspace{-0.5cm}

The autocorrelation time was monitored across the simulations. The observed upper bound on the thermalization time was $1500$ Monte Carlo sweeps for the parameters studied. In practice, $10^4$ sweeps were chalked up as a thermalization period and dropped from the  beginning of each Monte Carlo time series. The autocorrelation times in the thermalized time series were at most $200$ sweeps for the parameters studied and measurements were taken at intervals of $10^3$ sweeps to ensure that they were sufficiently decorrelated. For error estimation all error bars were computed using the jackknife method.

\vspace{-0.5cm}
\subsection{Overview}

\vspace{-0.2cm}

Let us step back and summarize what we have seen so far. As emphasized in the introduction, we have a lattice model with an Abelian gauge invariance of a non-standard variety. Charges are conserved but isolated charges are not present in the theory. Instead the simplest charge sector consists of four charges that can be moved around in pairs respecting the conservation of charge in each plane of the lattice.  The theory has no Lorentz invariance in the continuum limit nor even continuous rotational invariance and is instead tailored to exotic condensed matter applications stemming from fracton models such as the X-cube model \cite{Vijay_2016}. 

The rest of this article is devoted to an exploration of the quantum phases obtained by tuning the matter couplings. In this context it will be illuminating to refer to the behavior of the ordinary lattice Abelian Higgs model in $(3+1)D$ with lattice action
\begin{align}
S_{\rm ABH-Gauge} = -\beta\sum_{\tau=1}^N  \sum_{\vec{r}}& \cos( \Delta_i A_{j}(\vec{r})-\Delta_j A_{i}(\vec{r}) \\
&+ \cos (\Delta_i A_0^\tau(\vec{r}) -\Delta_0 A_{i}^\tau (\vec{r})) \nonumber 
	\label{eq:abhgauge}
\end{align}
and
\begin{align}
S_{\rm ABH-Higgs} = -\kappa\sum_{\tau=1}^N  \sum_{\vec{r}}&  \cos(qA_0^\tau(\vec{r})+\Delta_0 \theta^\tau (\vec{r}))\\
&+ \sum_{i<j} \cos(qA_{i}^\tau(\vec{r})+\Delta_i \theta^\tau(\vec{r}))\nonumber.
	\label{eq:abhhiggs}
\end{align}
Here there is $A_i(\vec{r})\rightarrow A_i(\vec{r}) + \Delta_i \Omega(\vec{r})$, $\theta(\vec{r})\rightarrow \theta(\vec{r}) - \Omega(\vec{r})$ gauge invariance.  
While this model is well known we briefly review the salient features. In the pure gauge theory $\kappa=0$, in $3+1$D the large $\beta$ phase is in the so-called Coulomb phase with a gapless, linearly dispersive photon and gapped magnetic monopole excitations \cite{fradkinshenker1979, guth1980}. In this phase, electric charges are deconfined and the naive continuum field theory of scalar Maxwell electrodynamics describes the long wavelength behavior of the theory. In contrast, in the small $\beta$ phase the photon is gapped out. Separating electric charges produces an electric flux string between them and hence a linear confining potential. Relatedly Wilson line operators obey an area law in the confined phase and a perimeter law in the deconfined law. The Coulomb phase is stable to switching on finite $\kappa$ but there is a first order phase boundary at some $\kappa(\beta)$ at which the gapped charges condense into a Higgs phase and the photon is gapped out. From the pure gauge confined phase, screening by charges means that the area law is replaced by a perimeter law for infinitesimal coupling $\kappa$. Osterwalder and Seiler \cite{osterwalderseiler} and then Fradkin and Shenker \cite{fradkinshenker1979} showed that there is a continuous region for $q=1$ where Green's functions are analytic that overlaps with parts of the Higgs phase and into the confining regime. These are therefore the same thermodynamic phase. When $q=2$ instead, the Higgs and confined regimes are different phases. The phase boundary between them separates a regime with deconfined $\mathbb{Z}_2$ charges (Higgs) from a phase where all charges are gapped. 

As we shall see the phase diagram for the A tensor gauge theory differs from its relative $-$ the $3+1$ dimensional Abelian-Higgs model $-$ in several important respects.  

\vspace{-0.5cm}
\section{Limiting Cases}

\vspace{-0.5cm}

To gain an understanding of the nature of the phases of the lattice gauge theory it is useful, first, to consider limiting cases where one coupling is either sent to zero or to infinity corresponding to the four bounding edges of the phase diagram.  

\vspace{-0.8cm}

\subsection{Switching off the gauge dynamics}

\vspace{-0.7cm}

The simplest case of the four is the one where we switch off the gauge coupling $\beta\rightarrow 0$ entirely so that the action only has the $S_{\rm Higgs}$ term. In the Hamiltonian language this is equivalent to sending $g_m/g_e\rightarrow \infty$. We may go to the unitary gauge $\theta^\tau (\vec{r})=0$ to reveal that the model consists only of decoupled plaquettes. The theory is therefore trivial in this limit. 

\vspace{-0.8cm}

\subsection{Infinite matter coupling} \label{subsection:InfMatter}

\vspace{-0.7cm}

If instead we take the matter-gauge coupling to infinity $\kappa\rightarrow\infty$, the matter fields and gauge fields are locked to one another with the constraint $pA_{ij}^\tau(\vec{r})+\Delta_i \Delta_j \theta^\tau(\vec{r})=0$ with a similar expression for the temporal links. Once again it is easiest to see what happens in the unitary gauge. There one finds the constraint $A_{ij}^\tau(\vec{r}) = 2\pi n_{ij}^\tau(\vec{r})/q$ where $n_{ij}^\tau(\vec{r})$ is integer valued. In the following we consider two cases $q=1$ and $q=2$. In the former case, only gauge equivalent configurations appear in the theory for any $\beta$ so again the theory is trivial. However, in the $q=2$ case, the plaquette and link fields are $\mathbf{Z}_2$ gauge fields and the theory is a pure fractonic $\mathbf{Z}_2$ gauge theory. 

We examine the $q=2$ case from the perspective of the Hamiltonian. The analog of the $\kappa\rightarrow \infty$ limit is then $MV \rightarrow \infty$. In this limit, the matter field ceases to be dynamical and we may gauge it away. With $V$ the only scale in the problem, the single plaquette ground states are those with $\vert A_{ij}(\vec{r}) = 0,\pi \rangle$ when $q=2$. Considering now the pure gauge Hamiltonian in this subspace we observe that the $L_{[ij]k}$ operators are diagonal in the single plaquette basis with eigenvalues $+ 1$ for $\vert A_{ij}(\vec{r}) = 0 \rangle$ and $-1$ for $\vert A_{ij}(\vec{r}) = \pi \rangle$. In other words, we may replace these operators with products of $\sigma^z$ Pauli operators on the elementary spatial tubes. Raising and lowering operators between these states modulo $2\pi$ are $\exp (\pm i \pi E_{ij}(\vec{r}))$. Within the large $V$ subspace, these act like $\sigma^x$ on plaquettes.  We may then introduce a term into the Hamiltonian that enforces the Gauss law dynamically. Since the matter fields have dropped out this takes the form
\be
U \sum_{\vec{r}} \prod_{\{ij\} \in \vec{r} }  (-)^{\Delta_i\Delta_j E_{ij}(\vec{r})} \sim U \sum_{\vec{r}} \prod_{\{ij\} \in \vec{r} }  \sigma^x_{ij}(\vec{r}) 
\ee
where the product runs over the $12$ plaquettes connected to site $\vec{r}$. The dual cubic lattice has sites at the center of cubes on the direct lattice. Mapping the model to the dual lattice leads to variables living on links instead of plaquettes. The resulting model is recognizably the X-cube model which has fractonic excitations \cite{Vijay_2016}. The observation that the X-cube model arises deep within the Higgs regime of the hollow rank-2 gauge theory coupled to $q=2$ scalar matter was reported in Refs.~\cite{Ma_2018,Bulmash_2018}. In particular, the X-cube mode arises when the matter dynamics has been frozen out and where the gauge electric field term has been set to zero. When $g_e$ is finite, the gauge fields become dynamical. Within the $2D$ local subspace such dynamics can be generated by a term of the form
\be
W \sum_{\vec{r},ij} \sigma^x_{ij}(\vec{r}). 
\ee
\vspace{-0.1cm}
We expect the fracton phase, whose parent Hamiltonian is the X-cube model, to be stable to the presence of the dynamical term up to some threshold. Switching once again to the action, we therefore expect a phase transition from the fractonic phase at large $\beta$ to a paramagnetic phase at small $\beta$. 

\vspace{-0.9cm}
\subsection{Infinite gauge coupling}
\vspace{-0.7cm}

In the limit $\beta\rightarrow\infty$, the gauge dynamics is frozen out and we may gauge away the $A_{ij}(\vec{r})$ fields leaving the matter action 
\be
S_{\rm \beta\rightarrow\infty} = -\sum_{\tau=1}^N  \sum_{\vec{r}} \kappa \cos(\Delta_0 \theta^\tau (\vec{r}))+ \sum_{i<j}\kappa \cos(\Delta_i \Delta_j \theta^\tau(\vec{r})).
	\label{eq:abhhiggs_betainfty}
\ee
In the ordinary compact $U(1)$ Abelian Higgs model, this limit leads to a lattice $XY$ model in 4D with exchange on links and with critical coupling $\kappa_c$. In the rank two case there is an XY degree of freedom on all sites but the temporal and spatial couplings are inequivalent. The latter is analogous to a ring exchange term and this difference results in a considerable enhancement of the symmetries. While the ordinary lattice XY model has a single global $U(1)$ symmetry this XY-plaquette model has subsystem symmetries. To see this select one plane in a spatial direction (for example indexed by position $\vec{r}_0$ and directions $ij=xy,xz$ or $yz$) and perform a global shift on this plane of $\phi\rightarrow \phi+\alpha$. The action is invariant under this shift. There are $L_x + L_y + L_z$ such planes but there are constraints coming from the fact that a shift on all fixed $xy$ planes simultaneously is equivalent to a shift on all $xz$ or all $yz$ planes. In total there are $L_x + L_y + L_z-2$ $U(1)$ subsystem symmetries. One expects spontaneous breaking of these symmetries at large $\kappa$ and a paramagnetic phase at small $\kappa$. 

The continuum field theory that is expected to describe the long wavelength behavior of this model is 
\be
\mathcal{L} = \frac{m}{2} \left( \partial_0\phi \right)^2 - \frac{1}{4\mu}  \left( \partial_i \partial_j\phi \right)^2.
\ee

\vspace{-0.5cm}

\begin{figure}[H]
    \centering
    \includegraphics[width=1\columnwidth]{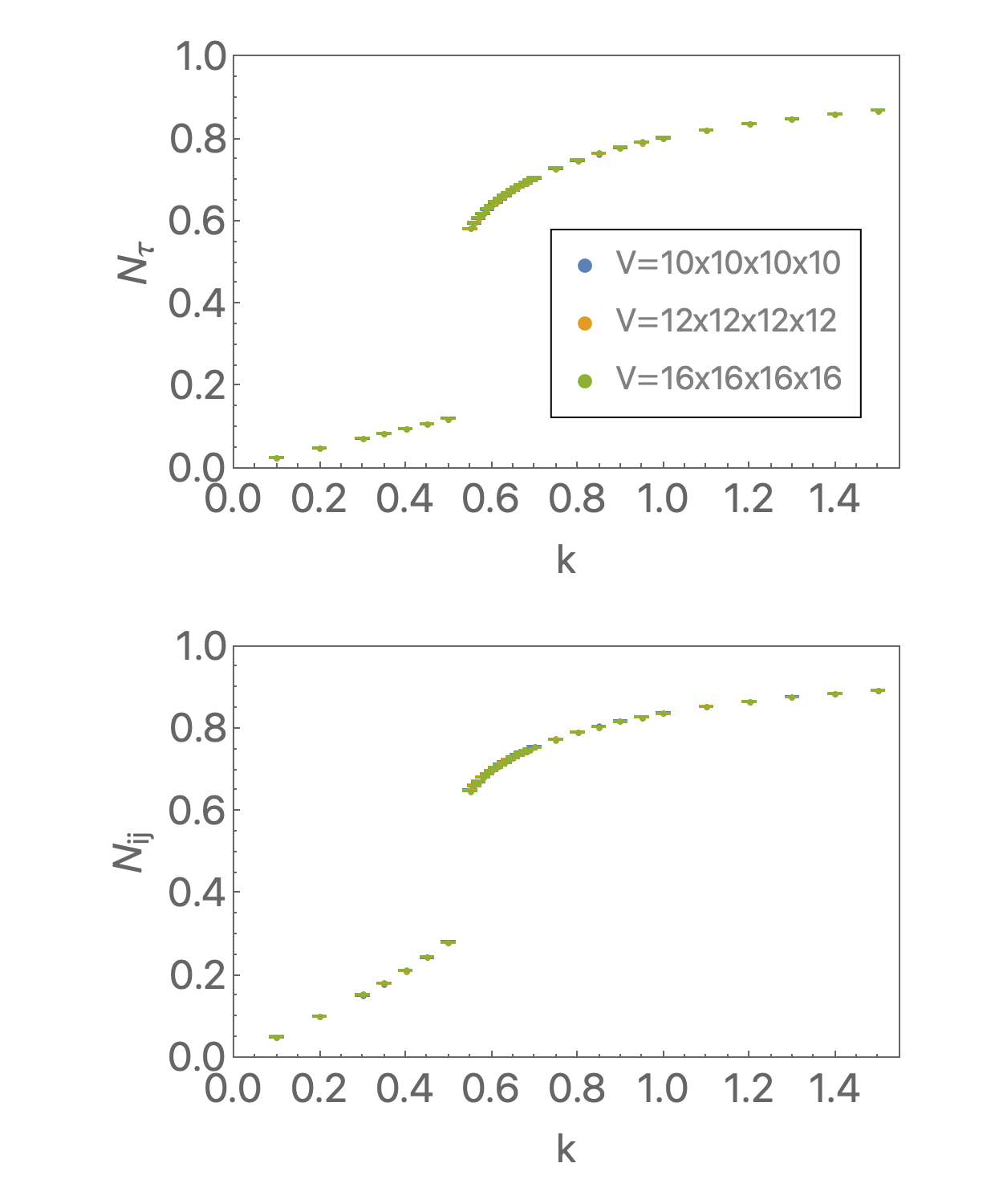}
    \caption{Figures showing the temporal-matter (top) and spatial-matter (bottom) expectation values for the XY plaquette model both showing a first order transition at $\kappa_c \approx 0.55$.}
    \label{fig:matteronly}
\end{figure}

We have carried out a Monte Carlo simulation of the model with action Eq.~\ref{eq:abhhiggs_betainfty} finding the expectation values of both the temporal links terms and the plaquette terms. Results are shown for a $10^4$ lattice in Fig.~\ref{fig:matteronly}. The simulation results provide evidence of a first order phase transition with $\kappa_c \approx 0.55$. 

\vspace{-1.5cm}
\subsection{Pure gauge theory at strong coupling}
\vspace{-0.5cm}

The final limiting case, $\kappa=0$, is also the most involved. This is the one where the matter-gauge coupling is taken to zero leaving only the pure gauge theory of Eq.~\ref{eq:higgslatticeaction}. In this subsection, we explore the strong ($\beta\ll 1$) coupling limit leaving the naive weak coupling analysis to an Appendix ($\beta\gg 1$). This organization reflects a crucial feature of the phase diagram of this model namely that the naive weak coupling theory breaks down as a consequence of the proliferation of instantons. 

We begin the strong coupling analysis with the partition function in the zero matter coupling limit, 
\begin{equation}
	Z = \int \prod_{\vec{x},\mu, \tau} dA_{\mu}(\vec{x},\tau) e^{-S[A]}
\end{equation}
where $\mu \in \{ 12,13,23,0\}$ and $S[A]=S_{\rm ABH-Gauge}+S_{\rm ABH-Higgs}$ is the action derived in the previous section. Let us consider arbitrary combinations of the elementary gauge invariant objects shown in Fig.~\ref{fig:gauge_inv} that we call schematically $\exp( i E^\tau_{ij}(\vec{r}))$ for the time-space cube and $\exp ( i B^\tau_{[ij]k}(\vec{r}))$ for the spatial hollow cube. We wish to evaluate
\be
\left\langle \prod_{E\in \mathcal{C}_E} e^{iE}   \prod_{B \in \mathcal{C}_B} e^{iB}   \right\rangle.
\label{eq:vev}
\ee
To take the strong coupling limit, we may expand $\exp( -\beta S)$ in powers of $\beta$. Inspecting Eq.~\ref{eq:vev} term by term in this expansion, one sees that the integrals over the gauge fields vanish unless the observables to be computed cancel with the conjugate term coming from products of 
\be
\frac{1}{2} \left[ e^{ i ( \Delta_i A_{jk}(\vec{r})-\Delta_j A_{ki}(\vec{r}))} + e^{ - i ( \Delta_i A_{jk}(\vec{r})-\Delta_j A_{ki}(\vec{r}) ) } \right]
\ee
or
\be
\frac{1}{2} \left[ e^{ i ( \Delta_i \Delta_j A_0^\tau(\vec{r}) -\Delta_0 A_{ij}^\tau (\vec{r})) } + e^{ -i ( \Delta_i \Delta_j A_0^\tau(\vec{r}) -\Delta_0 A_{ij}^\tau (\vec{r})) } \right]
\ee
in the action. If set $\mathcal{C}_E$ contains $N_E$ elements and $\mathcal{C}_B$ contains $N_B$ elements we find that
\be
\left\langle \prod_{E\in \mathcal{C}_E} e^{iE}   \prod_{B \in \mathcal{C}_B} e^{iB}   \right\rangle = \exp\left[ (N_E + N_B) \ln \left( \frac{\beta}{2} \right) \right].
\label{eq:vev2}
\ee

\vspace{-0.5cm}
It is instructive to consider some special cases of this result. One case is the expectation values of elementary plaquettes. At strong coupling the general result reveals that

\vspace{-1cm}
\be
\langle L_{ij\tau} \rangle_{\rm SC} = \langle L_{[ij]k} \rangle_{\rm SC} = \frac{\beta}{2}.
\label{eq:LijSC}
\ee
A second useful observable is obtained by taking the product of time-space cubes along the temporal direction. This is the hollow rank$-2$ analog of the Polyakov loop and we call it the time tube $L_{ij}$:

\vspace{-0.5cm}
\be
L_{ij}(\vec{r}) =\prod_{n=0}^{N_t-1}L_{ij\tau}(\vec{r}+n\vec{e}_\tau)= \exp (-i\sum_{n}^{N_t-1}\Delta_i \Delta_j A^\tau(\vec{r}+n\vec{e}_\tau))
\ee

\vspace{-0.5cm}

One can think of this as a world line of two pairs of opposite charges separated by one link and in the strong coupling limit we find
\be
\langle L_{ij}(\vec{r}) \rangle_{\rm SC} = \exp\left(  N_\tau \ln\left( \frac{\beta}{2} \right) \right).
\ee
Finally we generalize this result to the case where the charge world lines are placed on the corners of a spatial plaquette of length $R_i$ in one direction and $R_j$ in another direction measured in lattice spacings. This gives a handle on the potential involved in separating the charges. 
\be
\langle L_{ij}(\vec{r};R_i, R_j) \rangle_{\rm SC} = \exp\left(  N_\tau R_i R_j \ln\left( \frac{\beta}{2} \right) \right).
\ee
This shows that there is a linear confining potential between pairs of dipoles and a quadratic potential for separating pairs of dipoles into the constituent charges. 

\par Even more generally, consider a large hollow spatial cube with edge length $L$. Then the expectation of this in the small $\beta$ phase goes like $\exp(-L^3 \ln (\beta/2))$. Similarly for a time tube with both temporal and spatial extents being equal to $L$. 

\par To probe the potential between charges it is necessary to extend the time-cubes in the temporal direction, the time tube. But this object decays exponentially with the temporal span of the system, making it difficult to probe accurately using numerical methods. To verify the validity of strong coupling expansion it is useful the look at a simpler object, the extended time-cube $L_{ij\tau}^r(\vec{r})$, which is analogous to the extended time-tube but without the temporal elongation.

\vspace{-0.7cm}

\be
\langle L_{ij\tau}^r \rangle_{\rm SC} = \exp\left( r^2 \ln\left( \frac{\beta}{2} \right) \right).
\ee

\vspace{-0.2cm}

\begin{figure}[H]
\begin{center}
  \includegraphics[width=0.9\columnwidth]{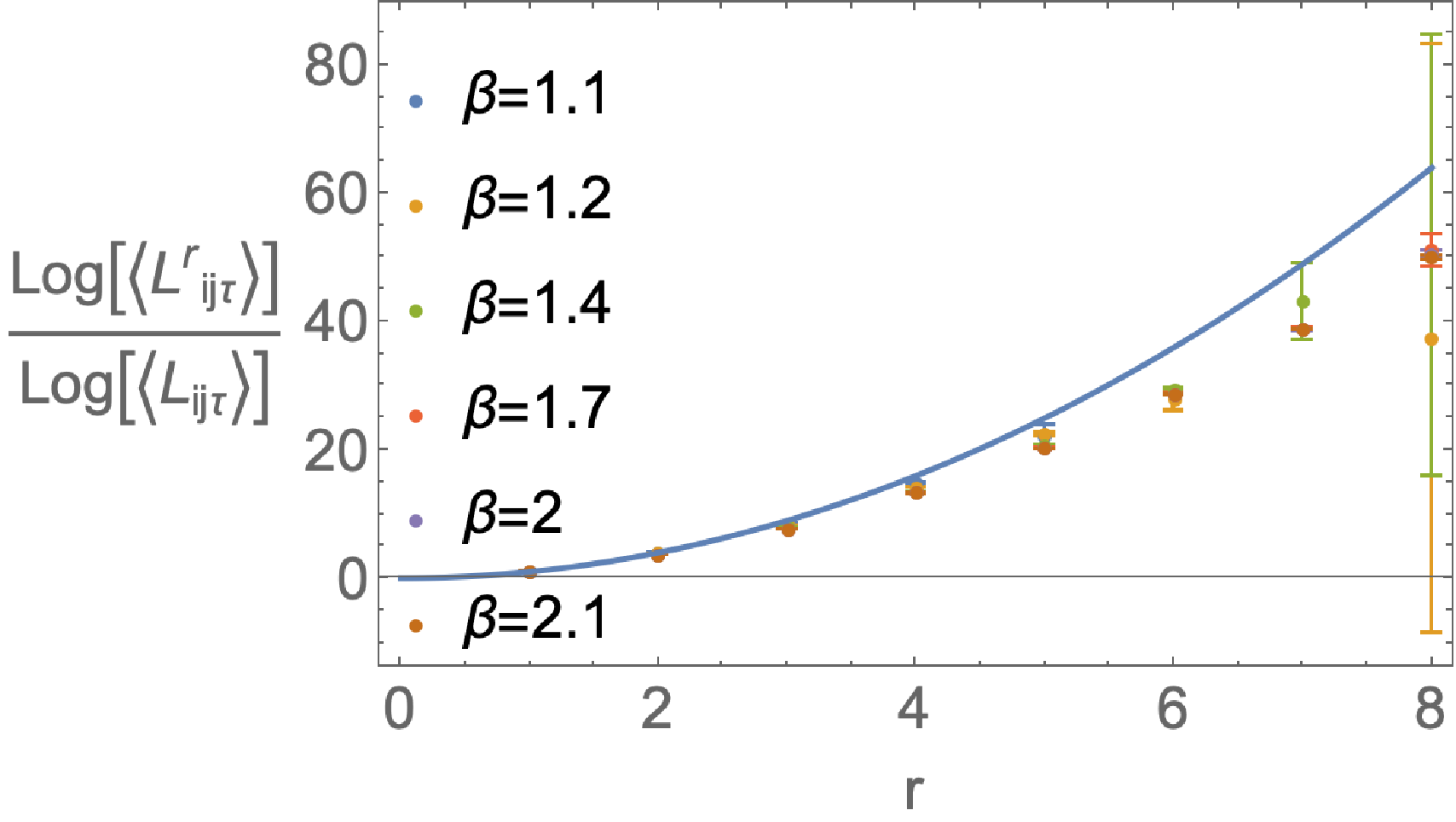}
  \end{center}
  \caption{Normalized logarithm of the extended time cubes for a system size $V=24x24x24$. Next to the data we plot $r^2$ for comparison. }
\end{figure}

\vspace{-0.7cm}
 
It is worth comparing these results to the behavior of ordinary $U(1)$ lattice gauge theory in 4D at strong coupling. As is well known, the theory at strong coupling lives in a confined phase with an area law Wilson loop and a linear potential between pairs of charges. In the rank$-2$ theory, rather than pairs of charges pairs of dipoles are confined in this limit and  the analog of the area law Wilson line comes from a loop that separates pairs of dipoles in space. The Polyakov loops are quantitatively identical in the rank-$2$ and ordinary $U(1)$ theories. However, the form of the gauge invariant objects in the action indicates the natural analogs of Wilson loops in the higher rank theory  occupy a 3-volume. At strong coupling these then obey a volume law. 

\vspace{-0.2cm}

\begin{figure}[H]
    \centering
    \includegraphics[width=0.9\columnwidth]{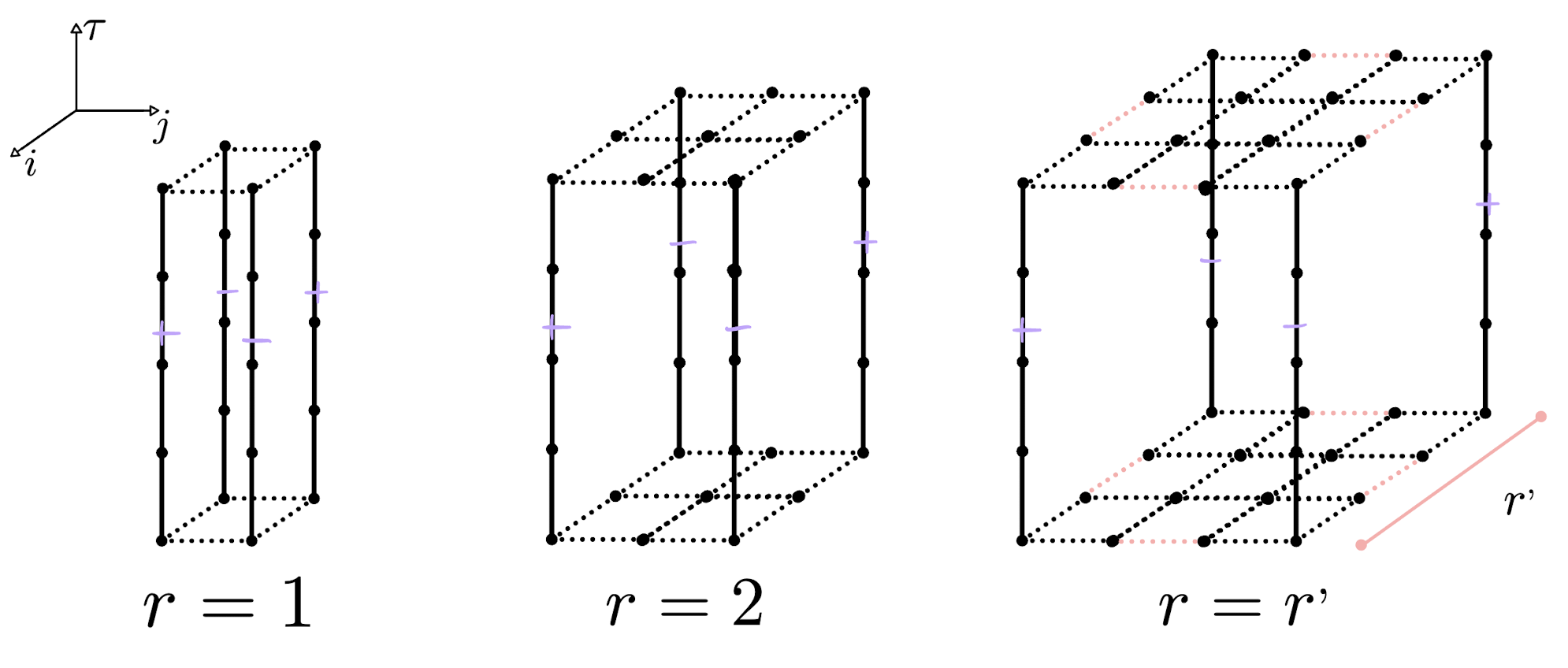}
    \caption{Figure showing expanded time tubes with spatial extent $r'$ used to measure the quadrupolar potential. }
    \label{fig:potential}
\end{figure}

\vspace{-0.5cm}

In Appendix~\ref{sec:weak_coupling_pinch_points}, we discuss the naive weak coupling limit where the pure gauge theory on the lattice maps to a tensor gauge field theory. We summarize the main results here and compare them to the strong coupling case. For example, we find at weak coupling
\begin{equation}
\langle L_{ij\tau} \rangle_{\rm WC} = \exp \left( - C/\beta \right)
\end{equation}
to be contrasted with Eq.~\ref{eq:LijSC} at strong coupling. The constant $C$ is a non-universal constant.

For $L_{ij}$ wrapping around in the temporal direction with spatial extent $r$, the weak coupling result is
 \begin{equation}
\left\langle L_{ij}(\vec{r};r) \right\rangle_{\rm WC} =\exp \left(  - \frac{r N_t C'}{\beta}  \right)
 \end{equation}
for some constant $C'$ and the strong coupling case ($\kappa=0$) is 
\vspace{-0.3cm}
 \begin{equation}
\left\langle L_{ij}(\vec{r};r) \right\rangle_{\rm SC} =\exp \left(  r^2 N_t \log\left( \frac{\beta}{2} \right)  \right).
 \end{equation}
We note that switching on an infinitesimal $\kappa$ around the large $\beta$ phase leads to 
 \begin{equation}
\left\langle L_{ij}(\vec{r};r) \right\rangle_{{\rm SC},\kappa\neq 0} =\exp \left( -\zeta \left( 2 r^2 + 4 N_t \right)  \right).
 \end{equation}
This is analogous to the effect of small $\kappa$ on the strong coupling phase of ordinary U($1$) gauge theory where an area law for Wilson lines ($\kappa=0$) is replaced by a perimeter law ($\kappa\neq 0$).
 \vspace{-0.3cm}
 \par As in the standard $U(1)$ gauge theory, the rank two theory has a softer potential for the separation of pairs of dipoles in the weak coupling limit than in the strong coupling case where the $r^2$ dependence is analogous to area law behavior. As the distance scaling is qualitatively different in the two limits they cannot be interpolated by varying $\beta$. We would therefore expect a phase transition to separate the two regimes if they were stable phases. As we show in the next section, the strong coupling regime survives to large $\beta$ as a result of non-perturbative physics.  

\vspace{-1cm}
\section{Pure Gauge Theory: Duality, Instantons and Confinement}
\vspace{-0.5cm}

A classic result in theoretical physics is the Polyakov confinement mechanism in three dimensional pure U($1$) lattice gauge theory where instantons proliferate in the weakly coupled gauge theory and disorder the gauge field leading to a mass gap. In the analog of this physics in four dimensional pure gauge theory, one finds that magnetic monopole worldlines proliferate only below some finite coupling and there is a finite $\beta$ phase transition. 
 \vspace{-0.3cm}
\par It turns out that the hollow rank $2$ gauge theory in four dimensions behaves similarly to the three dimensional U($1$) gauge theory. Although proliferation of instantons in this theory was observed to occur in the very early literature on this topic \cite{xucenke2008}, we have revisited the dual description of the model to show explicitly that the instanton lead to confinement at large $\beta$ in the sense that strong coupling and dual weak coupling calculations coincide. The absence of a phase transition is then naturally to be expected and we explore this issue numerically in the next section. 
 \vspace{-0.3cm}
\par The perspective in this section connects to the work of Shao and Seiberg that discusses variants of continuum higher rank field theories and their duals \cite{Seiberg_2020,Gorantla_2021,Seiberg2023}. In particular, our work refers to the lattice version of the $\tilde{A}$ and $\hat{\phi}$ theories discussed in Ref~\cite{Seiberg_2020}.
 \vspace{-0.3cm}
\par We take a cubic lattice in four dimensions with sites $\vec{r}$ and directions labelled $\tau,x,y,z$. We introduce elementary fields $A_{ij}(\vec{r})$ defined to be symmetric in their indices $i,j=x,y,z$ {\it but with $i\neq j$ }  and living on plaquettes and $A_\tau(\vec{r})$ living on temporal links. These are gauge fields transforming as
 \vspace{-0.3cm}
\begin{align}
A_\tau \rightarrow A_\tau + \Delta_\tau \alpha \\ 
A_{ij} \equiv A_{ij} + \Delta_ i \Delta_ j  \alpha. 
\end{align}
 \vspace{-0.1cm}
We further define gauge invariant tensor electric and magnetic fields
\begin{align}
E_{ij} \equiv \Delta_\tau A_{ij} - \Delta_i \Delta_j A_\tau \\ 
B_{[ij]k } \equiv \Delta_i  A_{jk} - \Delta_ j A_{ik}
\end{align}
where $E_{ij}$ is symmetric in its indices while $B_{[ij]k}$ is antisymmetric in the $i,j$ indices and $k\neq i,j$. These definitions imply that
\vspace{-0.2cm}
\beq
B_{[xy]z} + B_{[yz]x} + B_{[zx]y} =0. 
\label{eq:bconstraint}
\eeq
\vspace{-0.6cm}

The partition function for the theory is
\vspace{-0.3cm}
\beq
Z=  \int DA \exp \left(  -\beta \sum_{r,i<j} \cos \left( B_{[ij]k}(\vec{r}) \right)+\cos \left( E_{ij}(\vec{r}) \right) \right).
\eeq

\vspace{-0.4cm}

Here the $A$ fields are defined from $-\pi$ to $\pi$. Without worrying about the range of $A$ we note that the electric and magnetic fields satisfy the analog of a Bianchi identity
\beq
\Delta_\tau B_{[ij]k} = \Delta_i E_{jk} - \Delta_j E_{ik}.
\eeq

\vspace{-0.5cm}

The behavior of the theory in the large $\beta$ limit is not as in the naive continuum limit. Instead one must consider carefully the effects of the compactness of the gauge fields. To make these effects more apparent we first ``Villainize" the action. The general idea is to make use of the identity:
\vspace{-0.2cm}
\beq
\exp \left( -\beta (1- \cos \theta) \right) = \sum_{m=-\infty}^{\infty} e^{i m\theta} e^{-\beta} I_{m}(\beta)
\eeq
where the modified Bessel function approximates to $e^{\beta} e^{-m^2/4\beta}/\sqrt{2\pi\beta}$ at large $\beta$ or weak coupling. Note that this explores the regime where there is a naive continuum limit of small $\theta$ fluctuations in the action. 

The analogs of integers $m$ for the hollow rank two theory are $n_{[ij]k}$ for $B_{[ij]k}$ and $n_{\tau ij}$ for $E_{ij}$. Use of the above identity in the action leads to the interpretation of the gauge potential as Lagrange multipliers that we integrate out to get constraint equations. These are
\begin{align}
\Delta_\tau n_{\tau yz} + \Delta_x \left( n_{[xy]z} + n_{[xz]y}  \right) & = 0 \\
\Delta_\tau n_{\tau xz} + \Delta_y \left( n_{[yz]x} + n_{[yx]z}  \right) & = 0 \\
\Delta_\tau n_{\tau xy} + \Delta_z \left( n_{[zx]y} + n_{[zy]x}  \right) & = 0 
\end{align}
\vspace{-0.3cm}
 and
 \vspace{-0.3cm}

 \vspace{-0.5cm}
 \beq
 \Delta_i \Delta_j n_{\tau ij} = 0.
 \eeq

 \vspace{-0.5cm}

We shall now introduce integer fields $\phi_{[ij]k}$ defined on the centers of spatial cubes. These allow one to solve the constraint equations as follows:
\vspace{-0.9cm}

\begin{align}
n_{[ij]k} & \equiv  -\Delta_\tau \phi_{[ij]k} \\
n_{\tau ij} & \equiv  \Delta_k \left( \phi_{[ki]j} + \phi_{[kj]i}  \right).
\end{align}

\vspace{-0.5cm}

Another way of proceeding is to naively write the cosine as $(\theta- 2\pi n)^2$ with a sum over $n$. Then we write the $n$ fields using the solution above plus integer versions of the electric and magnetic fields with an integer potential $h$. We may absorb the $h$ fields into the $A$ fields rendering them non-compact. In other words we define
\vspace{-0.2cm}
\begin{align}
\tilde{E}_{ij} & \equiv \Delta_\tau A_{ij} - \Delta_i \Delta_j A_\tau +  \Delta_\tau h_{ij} - \Delta_i \Delta_j h_\tau \\
\tilde{B}_{[ij]k} & \equiv  \Delta_i  A_{jk} - \Delta_ j A_{ik} +  \Delta_i  h_{jk} - \Delta_ j h_{ik} 
\end{align}
and then we evaluate the cross terms in the square that appear between the $\tilde{E}$, $\tilde{B}$ and the $\phi$ terms. It turns out that, after partial integration, the terms vanish as a consequence of the Bianchi identity. 
So what about the remaining terms in $\phi$? Using either approach, these are:
\beq
S_{\rm \phi} = \frac{1}{4\beta} \sum_{r,i<j} \left[ \left( \Delta_\tau \phi_{[ij]k} \right)^2 + \left( \Delta_k \left( \phi_{[ki]j} + \phi_{[kj]i}  \right) \right)^2 \right].
\eeq
The constraint on the components should be maintained when summing over configurations. 
\vspace{-0.3cm}
\par Expanding out the action and going to momentum gives the following kernel coupling $(\phi_{[xy]z}, \phi_{[yz]x}, \phi_{[zx]y})$:
\beq
\mathbf{M}(\mathbf{k}) = \left( \begin{array}{ccc} 
\omega^2 - k_x^2 - k_y^2 & k_y^2 & k_x^2 \\
k_y^2 & \omega^2 - k_y^2 - k_z^2 & k_z^2 \\
k_x^2 & k_z^2 &  \omega^2 - k_x^2 - k_z^2
 \end{array} \right)
\eeq
The denominator in the inverse of this kernel is
\beq
\omega^2 \left( \omega^4 - 2\omega^2 (k_x^2 + k_y^2 + k_z^2) + 3(k_x^2 k_y^2 + k_x^2 k_z^2 + k_y^2 k_z^2 ) \right)
\eeq
that gives the spectrum reported in the Seiberg {\it et al.} paper on their $\hat{\phi}$ theory \cite{Seiberg_2020}. 
\vspace{-0.3cm}
\par Now we go back and add a general source term to the action of the form
\beq
\sum_{\vec{r}, i<j} J_{ij} A_{ij} + J_\tau A_\tau
\eeq
so that now when the gauge fields are integrated out the constraint equations take the form
\begin{align}
\Delta_\tau n_{\tau yz} + \Delta_x \left( n_{[xy]z} + n_{[xz]y}  \right) & = J_{yz} \\
\Delta_\tau n_{\tau xz} + \Delta_y \left( n_{[yz]x} + n_{[yx]z}  \right) & = J_{xz} \\
\Delta_\tau n_{\tau xy} + \Delta_z \left( n_{[zx]y} + n_{[zy]x}  \right) & = J_{xy} 
\label{eq:nJ}
\end{align}
 and
 \beq
 \Delta_i \Delta_j n_{\tau ij} = J_\tau.
 \eeq
 summed over $i<j$.
\vspace{-0.3cm}
\par The analog of the continuity equation for this current is
\vspace{-0.5cm}
\beq
\Delta_\tau J_\tau = \Delta_i \Delta_j J_{ij}
\eeq
again summed over $i<j$. As may be verified by looking at Eqs.~\ref{eq:nJ}.
\vspace{-0.3cm}
\par In principle one may look for a general solution to these equations for the integer-valued $n$ and $\phi$. But we are interested in making contact with the strong coupling expansion. Therefore, we choose a particular form for the Wilson line mimicking the form we studied in that limit. This means choosing current lines $J_\tau(x,y,z,\tau) = \pm 1$ along four lines along the time direction centered at: 
\begin{align}
J_\tau = +1: x=-R/2, y=-R/2, z=0 \\
J_\tau = -1: x=+R/2, y=-R/2, z=0 \\
J_\tau = -1: x=-R/2, y=+R/2, z=0  \\
J_\tau = +1: x=+R/2, y=+R/2, z=0 
\label{eq:current_dist}
\end{align}
and zero everywhere else. This straightforwardly satisfies the continuity equation. 
\vspace{-0.3cm}
\par One may formally solve the equations as follows for this particular source term
\begin{align}
n_{\tau xy} & \equiv \Delta_z \left( \phi_{[zx]y} + \phi_{[zy]x}  \right) + \left( \Delta_x \Delta_y \right)^{-1} J_\tau \\
n_{\tau xz} & \equiv \Delta_y \left( \phi_{[yx]z} + \phi_{[yz]x}  \right)  \\
n_{\tau yz} & \equiv \Delta_x \left( \phi_{[xy]z} + \phi_{[xz]y}  \right)  \\
n_{[ij]k} & \equiv -\Delta_\tau \phi_{[ij]k} \hspace{1cm} {\rm for} \ ij = xy, xz, yz 
\end{align}

\vspace{-0.5cm}

Now let us recall the Poisson summation formula that introduces continuous fields into the problem
\beq
\sum_{p} f(p) = \sum_{q} \int d\psi f(\psi) e^{2\pi i q \psi }
\eeq
\vspace{-0.5cm}

Let us also neglect the constraint on the sums of the $\phi_{[ij]k}$. 
\vspace{-0.3cm}
\par We find, for the full dual action in the presence of currents:
\begin{align}
S_{\rm \psi, a}  = \frac{1}{4\beta} \sum_{r,i<j}&  \left( \Delta_\tau \psi_{[ij]k}  \right)^2 + \left( \Delta_k \left( \psi_{[ki]j} + \psi_{[kj]i}  \right) + I_{ij\tau} \right)^2  \\ 
& +  2\pi i a_{[ij]k} \psi_{[ij]k} \nonumber
\end{align}
where 
\beq
I_{ij\tau} = \delta_{i,x}\delta_{j,y}  \left( \Delta_x \Delta_y \right)^{-1} J_\tau.
\eeq

We introduce the three component field $(\psi_{[xy]z}, \psi_{[yz]x}, \psi_{[zx]y} )$ and defining the kernel 
\beq
\mathbf{M} \equiv  \left( \begin{array}{ccc} 
\Delta_\tau^2 + \Delta_x^2 + \Delta_y^2 & -\Delta_y^2 & -\Delta_x^2 \\
-\Delta_y^2 &\Delta_\tau^2 + \Delta_y^2 + \Delta_z^2  & -\Delta_z^2 \\
-\Delta_x^2 & -\Delta_z^2 & \Delta_\tau^2 + \Delta_x^2 + \Delta_z^2
 \end{array} \right).
\eeq
It is important to note that the order of indices does not follow a cyclic permutation. The choice recorded above with $ \psi_{[zx]y}$ instead of $ \psi_{[xz]y}$ is consistent with the sign structure in the double derivative operator $\mathbf{M}$. 
 \vspace{-0.3cm}
\par Now we may integrate out $\psi$ leaving an action for the $a_{[ij]k}$ and the currents. We find
\begin{align}
\beta  \left( 2\pi i a_{[ij]k}  + \frac{1}{2\beta} \mathcal{I}_{[ij]k} \right) \mathbf{M}^{-1}   \left( 2\pi i a_{[ij]k}  + \frac{1}{2\beta} \mathcal{I}_{[ij]k} \right)
\end{align}
where, for the particular charge configuration of interest:
 \vspace{-0.1cm}
\beq
\left(\mathcal{I}_{[xy]z}, \mathcal{I}_{[yz]x}, \mathcal{I}_{[zx]y} \right) = \left(0, \Delta_z  \left( \Delta_x \Delta_y \right)^{-1} J_\tau , -\Delta_z \left( \Delta_x \Delta_y \right)^{-1} J_\tau \right).
\eeq
To progress we regularize the potential writing 
\beq
\mathbf{M}^{-1}(\vec{r},\vec{r}') =  V_0 \delta_{\vec{r},\vec{r}'} + V(\vec{r}-\vec{r}')
\eeq
and we decouple the interaction between the $a_{[ij]k}$ fields by introducing new variables $\eta_{[ij]k}$.
 \vspace{-0.3cm}
\par We end up with the action
\begin{align}
S_{\rm \eta, a}  =  \sum_{r,i<j}& -\frac{1}{4\pi^2} \frac{1}{4\beta} \eta_{[ij]k} \mathbf{M} \eta_{[ij]k} +  i a_{[ij]k} \eta_{[ij]k} \\
&- 4\pi^2\beta  V_0  a_{[ij]k} a_{[ij]k} + 2 i\pi a_{[ij]k} \mathbf{M}^{-1} \mathcal{I}_{[ij]k} \nonumber 
\end{align}
\par Let us examine the part of the action that depends on $a_{[ij]k}$. These are integer valued fields that we must sum over:
\begin{align}
\sum_{a_{[ij]k} = -\infty}^{\infty} e^{ - 4\pi^2 \beta  V_0  a_{[ij]k} a_{[ij]k} +  i a_{[ij]k} \eta_{[ij]k} + 2 i\pi a_{[ij]k}\mathbf{M}^{-1} \mathcal{I}_{[ij]k} }
\end{align}
where clearly the regularization of the potential plays the role of suppressing the contribution from the $a_{[ij]k}$ when $\beta$ is sufficiently large (deep in the weak coupling regime). Essentially the monopole charges are then dilute and we may restrict the sum to $-1,0,1$. Doing this we find
\begin{align}
1 + & 2 \exp \left( - 4\pi^2\beta  V_0 \right)  \cos  \left(  \eta_{[ij]k} + 2\pi  \mathbf{M}^{-1} \mathcal{I}_{[ij]k}  \right) + \ldots \nonumber \\
& \approx \exp \left[ 2 \exp \left( - 4\pi^2\beta  V_0 \right)  \cos  \left(  \eta_{[ij]k} +  2\pi  \mathbf{M}^{-1} \mathcal{I}_{[ij]k} \right) \right].
\end{align}

\vspace{-0.3cm}

Overall we find for the partition function
\vspace{0.1cm}
\begin{align}
Z[L] =& \prod_{\vec{r}} \prod_{i<j \neq k} \int d\eta_{[ij]k} \exp \left[  \sum_{r,i<j} -\frac{1}{4\pi^2} \frac{1}{4\beta} \eta_{[ij]k} \mathbf{M} \eta_{[ij]k} \right. \nonumber \\ 
&\left. + 2 \exp \left( - 4\pi^2\beta  V_0 \right)  \cos  \left(  \eta_{[ij]k} + 2\pi  \mathbf{M}^{-1} \mathcal{I}_{[ij]k} \right)  \right] 
\label{eqn:weak_coupling_part_func}
\end{align}

\vspace{-0.5cm}

We may now compute the saddle point equations for the dynamical $\eta_{[ij]k}$ fields:
\beq
\mathbf{M}_{ab} \eta_b = - M^2 \sin\left(  \eta_a+ 2\pi  \mathbf{M}^{-1} \mathcal{I}_{[ij]k}  \right)
\eeq
where $a=[ij]k$. So there are three coupled differential equations here to solve in the presence of a source term.

\par We may solve these equations numerically without any further assumptions to understand the quantitative effective potential of the model at weak coupling. By introducing a fictional temperature $T$ we can use simulated annealing Monte Carlo to find a minimum for the action present in the exponent of \ref{eqn:weak_coupling_part_func}.
\hspace{-0.3cm}
\par Below we present the annealing results for a system with size $V=30\times 30\times 30$ and the current distribution from equations \ref{eq:current_dist}.

\onecolumngrid

\begin{figure}[H]
\begin{center}
  \includegraphics[width=0.9\columnwidth]{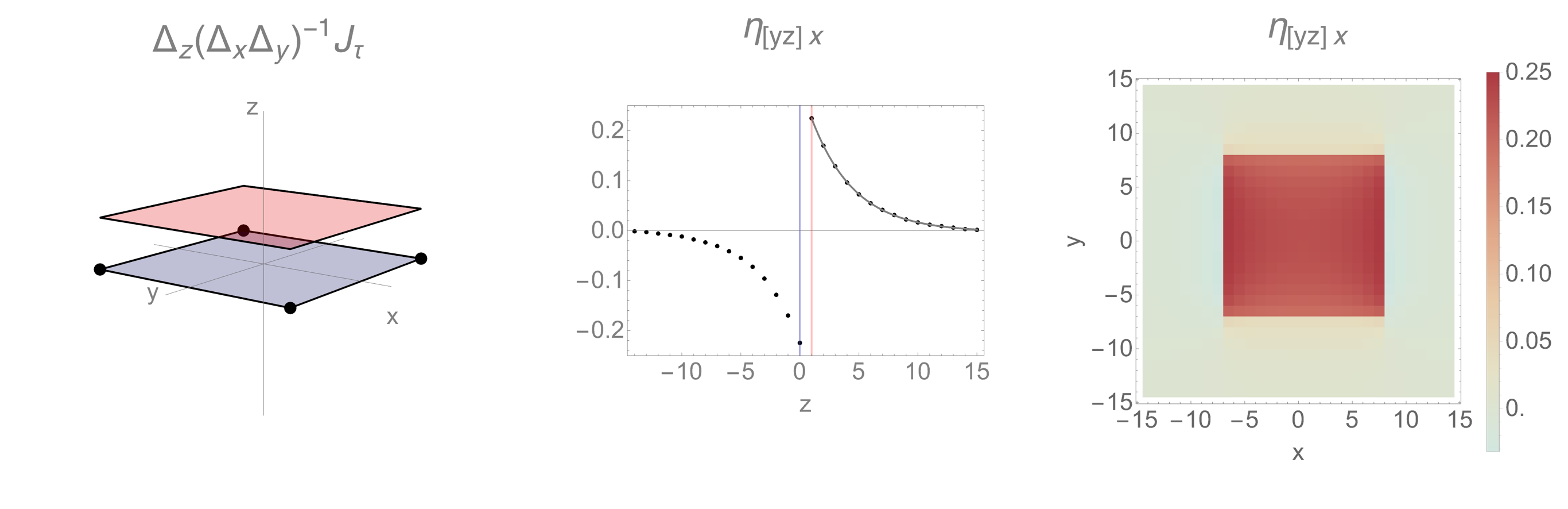}
  \caption{Right: Current configuration used. The red plane consists of positive charge and the blue plane of negative charge Center: $\eta_{[yz]x}$ field z-dependence for x and y inside the current configuration. The vertical lines correspond to the location of the charge planes of the corresponding color. Right: $\eta_{[yz]x}$ fields distribution for z=1}
  \label{fig:instantons}
  \end{center}
\end{figure}
\vspace{1cm}

\twocolumngrid

\par The results from the annealing process show that the $\eta_{[yz]x}$ fields are accumulating near the currents distribution. As the distance from the places increases the field magnitude decreases exponentially, as can be seen from the fit in the center plot of figure \ref{fig:instantons}. Outside of the planes the fields contributions are negligible. Similar results were obtained for the $\eta_{[zx]y}$, but with opposite sign. The contributions from $\eta_{[xy]z}$ are in general negligible when compared to the other fields. 

\par The accumulation of the fields over the current distribution seems to suggest an area law. To verify this, the ground state energy obtained from the annealing procedure was calculated for different charge distributions with varying linear separations $R$ between charges. 

\vspace{-0.1cm}

\begin{figure}[H]
\begin{center}
  \includegraphics[width=\columnwidth]{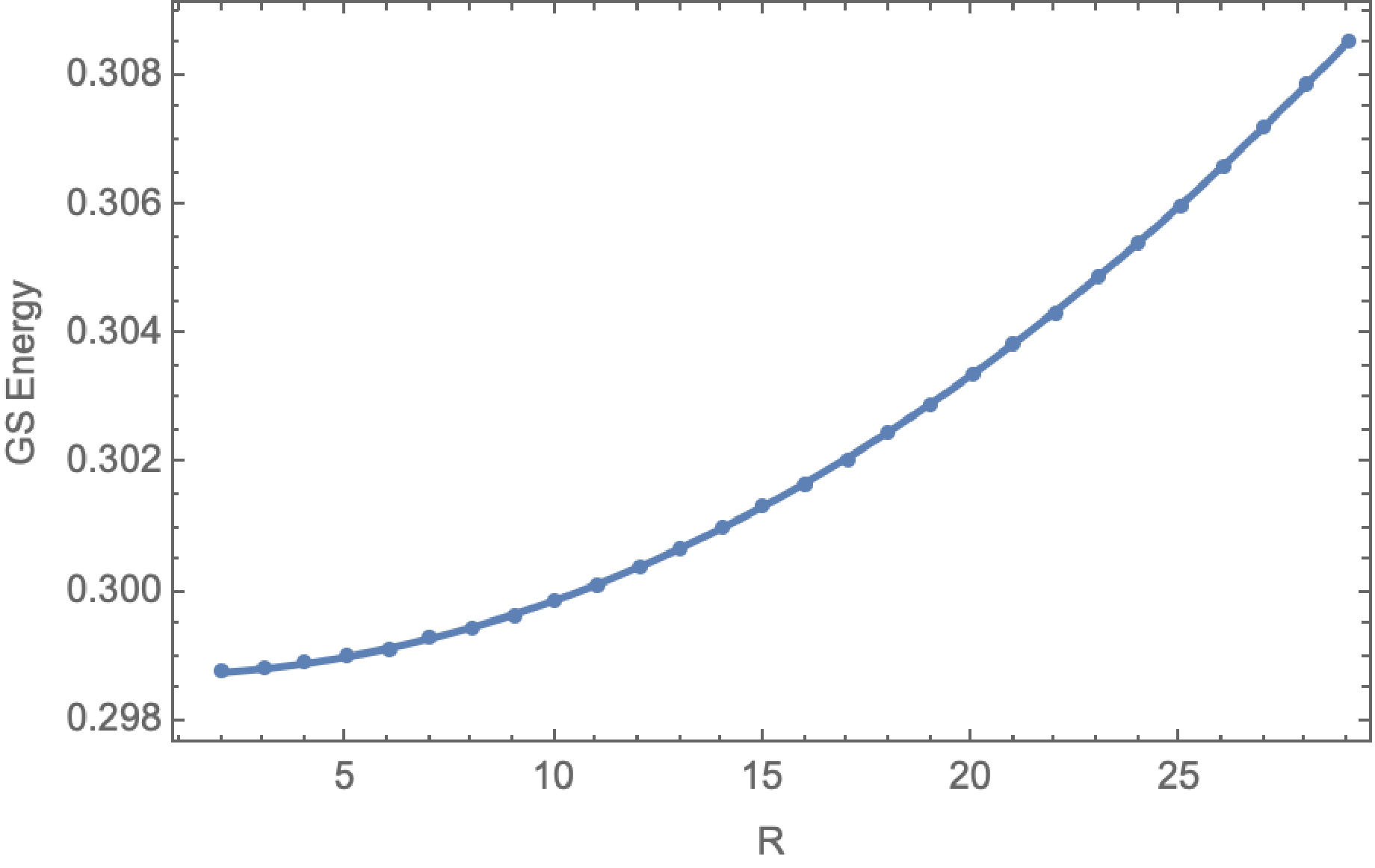}
  \end{center}
  \caption{Ground state energy dependence on the charge pair separation R. The data was fitted to $a+bx+cx^2$ and the resulting curve was plotted on top of the data. The fitting parameters obtained were $a=0.299$, $b=-5.02\times10^{-6}$ and $c=1.18\times10^{-5}$, which shows the linear term is negligible and the expression obtained is purely quadratic in nature.  }
\end{figure}

\vspace{-0.5cm}

\par By fitting the data we found that the ground state energy scales exactly like $R^2$, in agreement with the result found for the strong coupling regime.

\par This result implies that the point-like instanton solutions will not shield the strong potential from the strong-coupling regime and it will carry on for weak-coupling. Therefore there is no phase transition between these two regimes for the hollow rank-2 gauge theory.

\vspace{-0.5cm}

\section{Phase Diagram of the $A$ tensor gauge theory}

\begin{figure}[H]
    \centering
    \includegraphics[width=0.8\columnwidth]{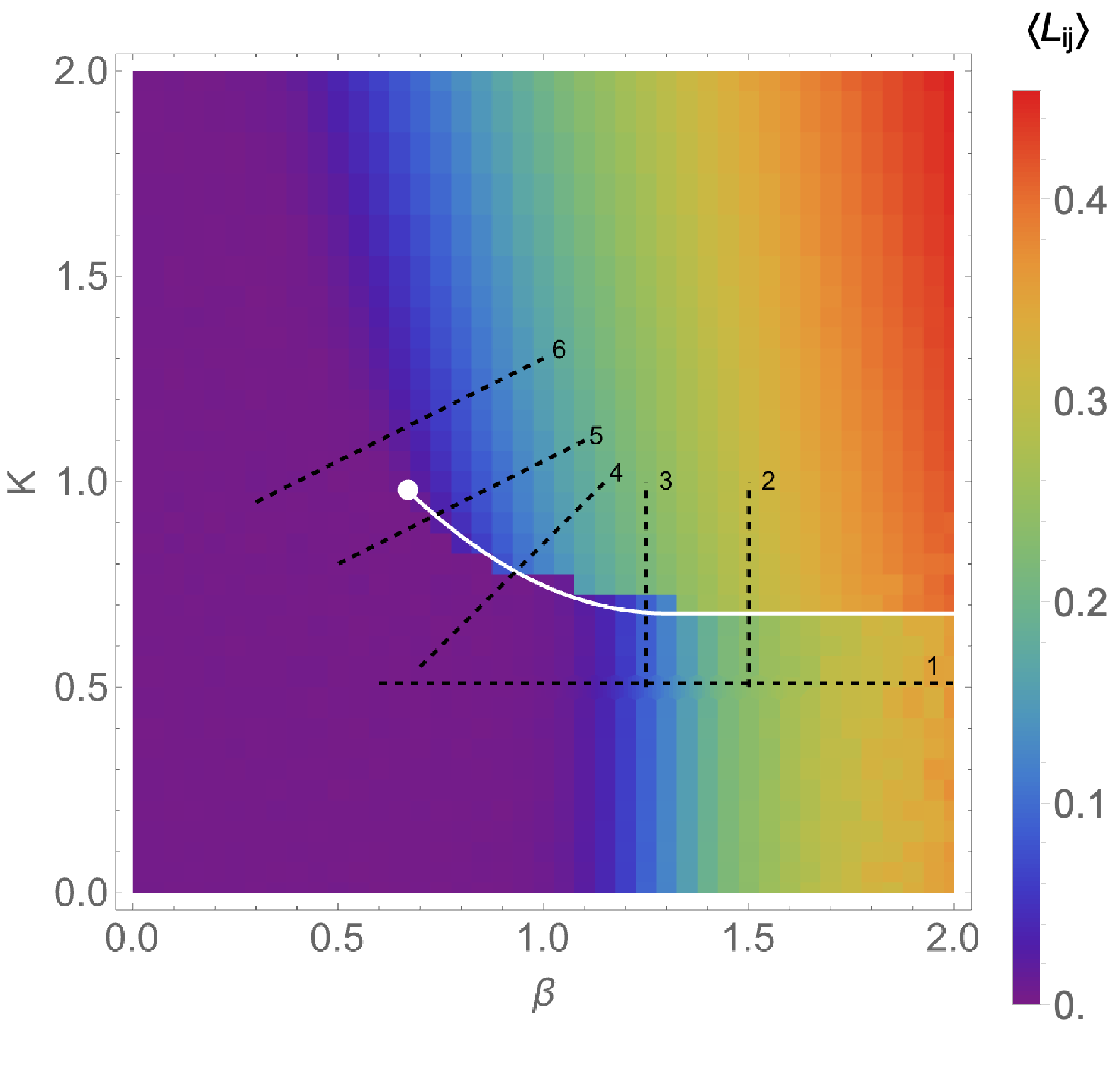}
    \caption{Phase diagram of the $4D$ charge $1$ Higgsed hollow rank 2 gauge theory parametrized by $\beta$ and $\kappa$.The order parameter plotted is the time tube $\langle L_{ij}\rangle$. The small $\kappa$, large $\beta$ phase is connecting to the other phase by horizontal and vertical phase boundaries. There is a first order phase boundary (white line) connecting to a critical endpoint (white point). The color gradient is the time tube expectation value. Dashed black lines indicate lines along which cuts are shown in figures below. This phase diagram was obtained from Monte Carlo simulations on a $16^4$ lattice. }
    \label{fig:TTcharge1}
\end{figure}

\begin{figure}[H]
     \centering
     \begin{subfigure}[b]{0.45\textwidth}
         \centering
         \includegraphics[width=\textwidth]{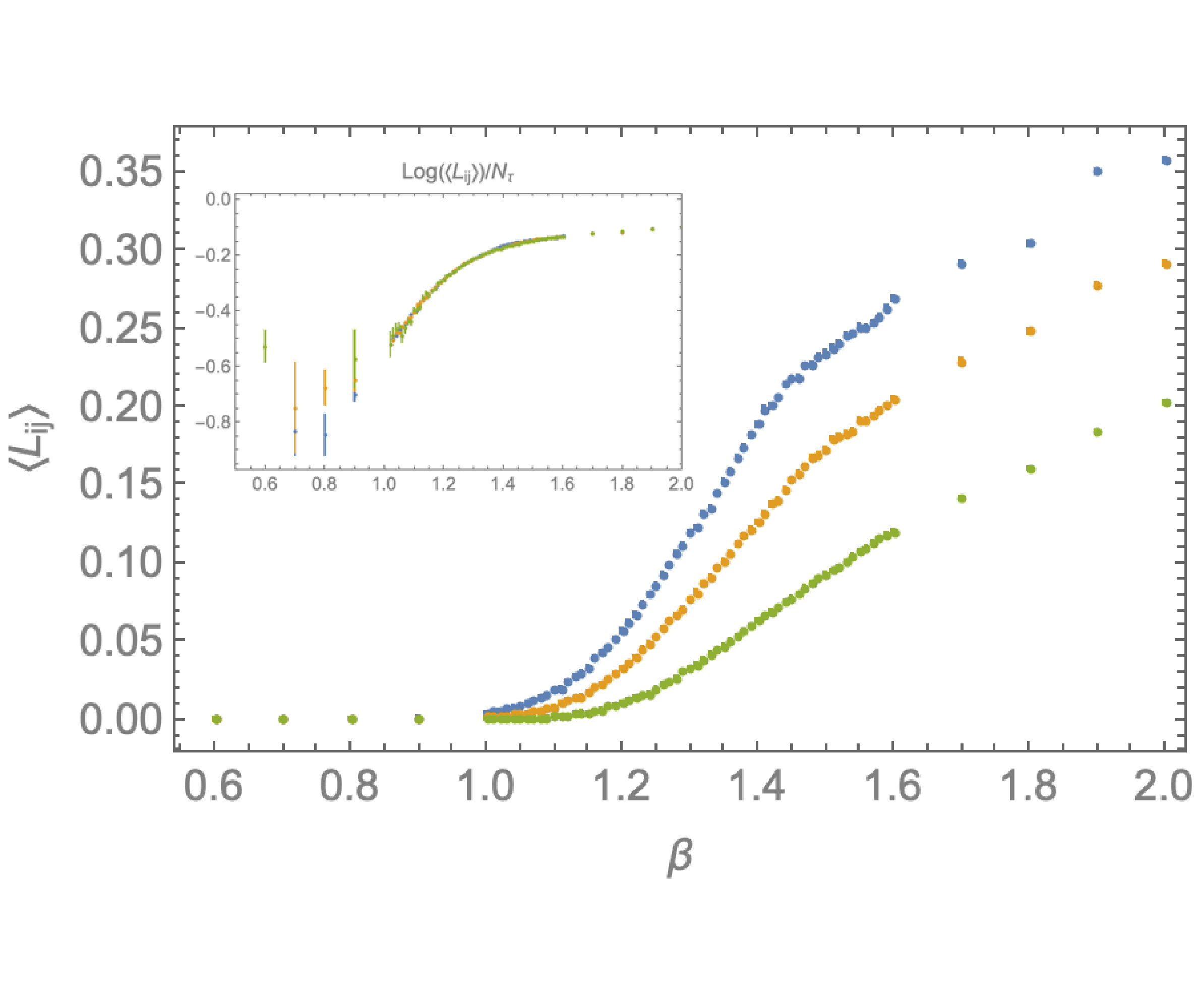}
         \caption{Line 1}
     \end{subfigure}
     \hfill
     \begin{subfigure}[b]{0.45\textwidth}
         \centering
         \includegraphics[width=\textwidth]{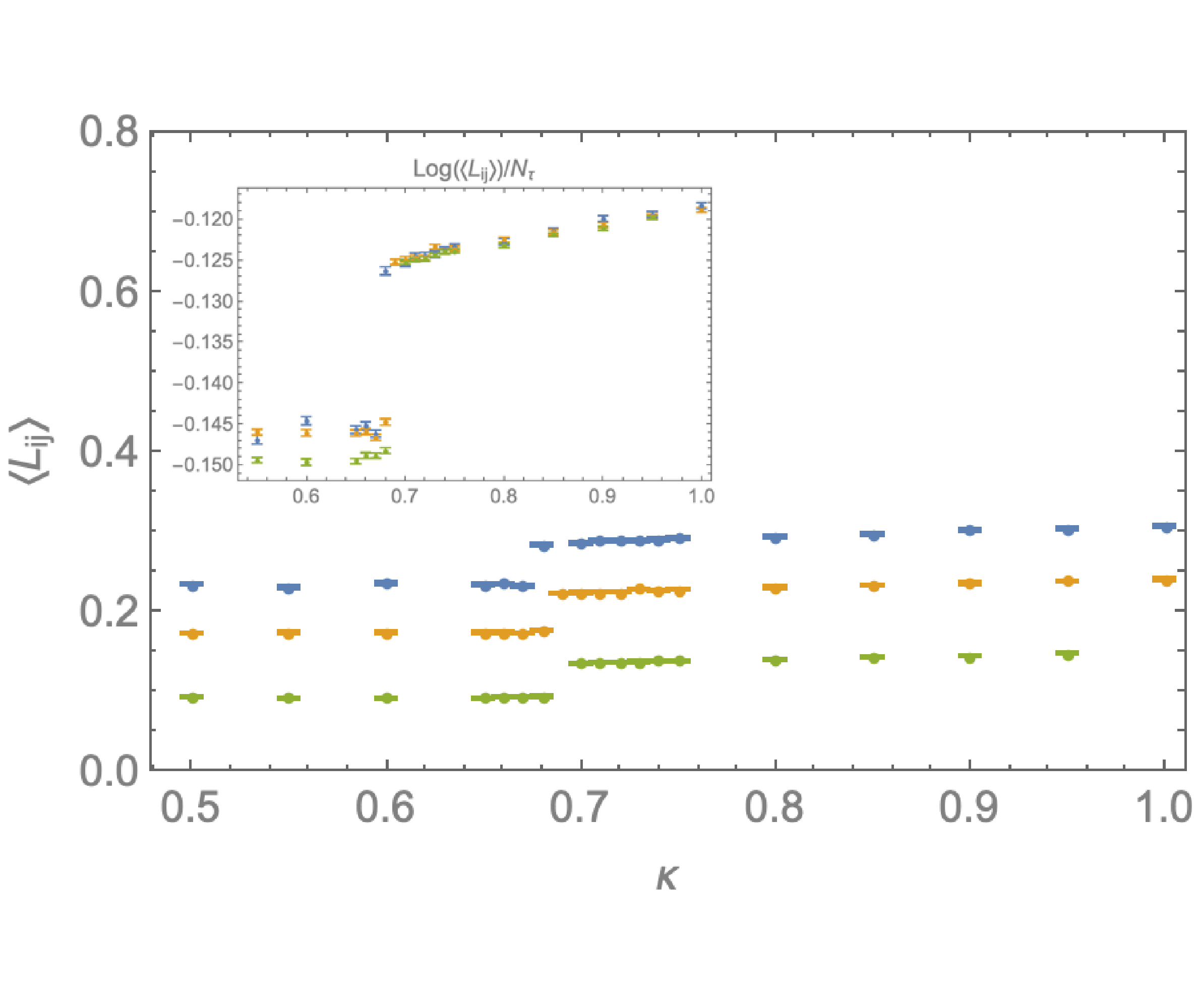}
         \caption{Line 2}
     \end{subfigure}
     \hfill
     \begin{subfigure}[b]{0.45\textwidth}
         \centering
         \includegraphics[width=\textwidth]{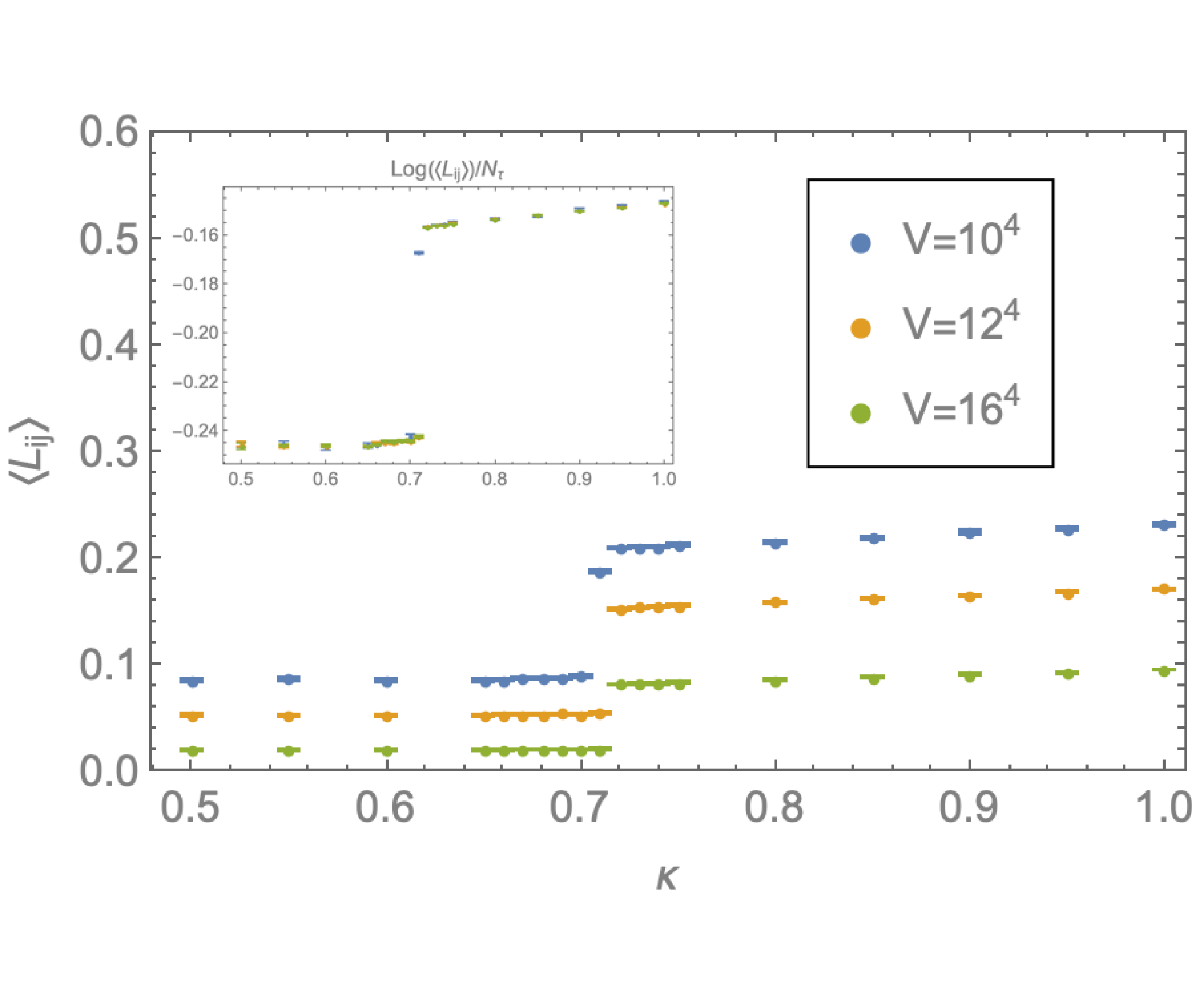}
         \caption{Line 3}
     \end{subfigure}
     \hfill
     \begin{subfigure}[b]{0.45\textwidth}
         \centering
         \includegraphics[width=\textwidth]{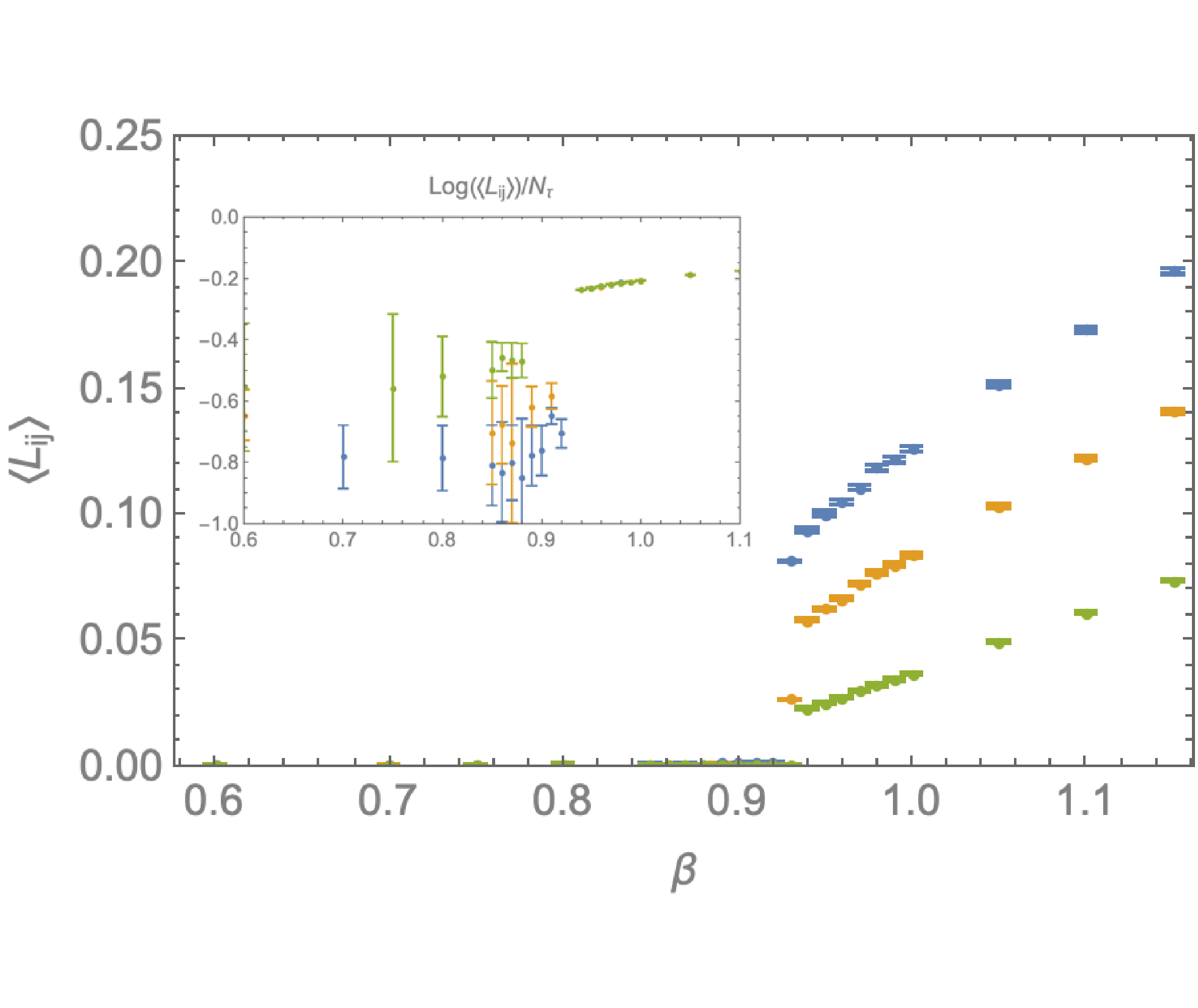}
         \caption{Line 4}
     \end{subfigure}
     \hfill
     \begin{subfigure}[b]{0.45\textwidth}
         \centering
         \includegraphics[width=\textwidth]{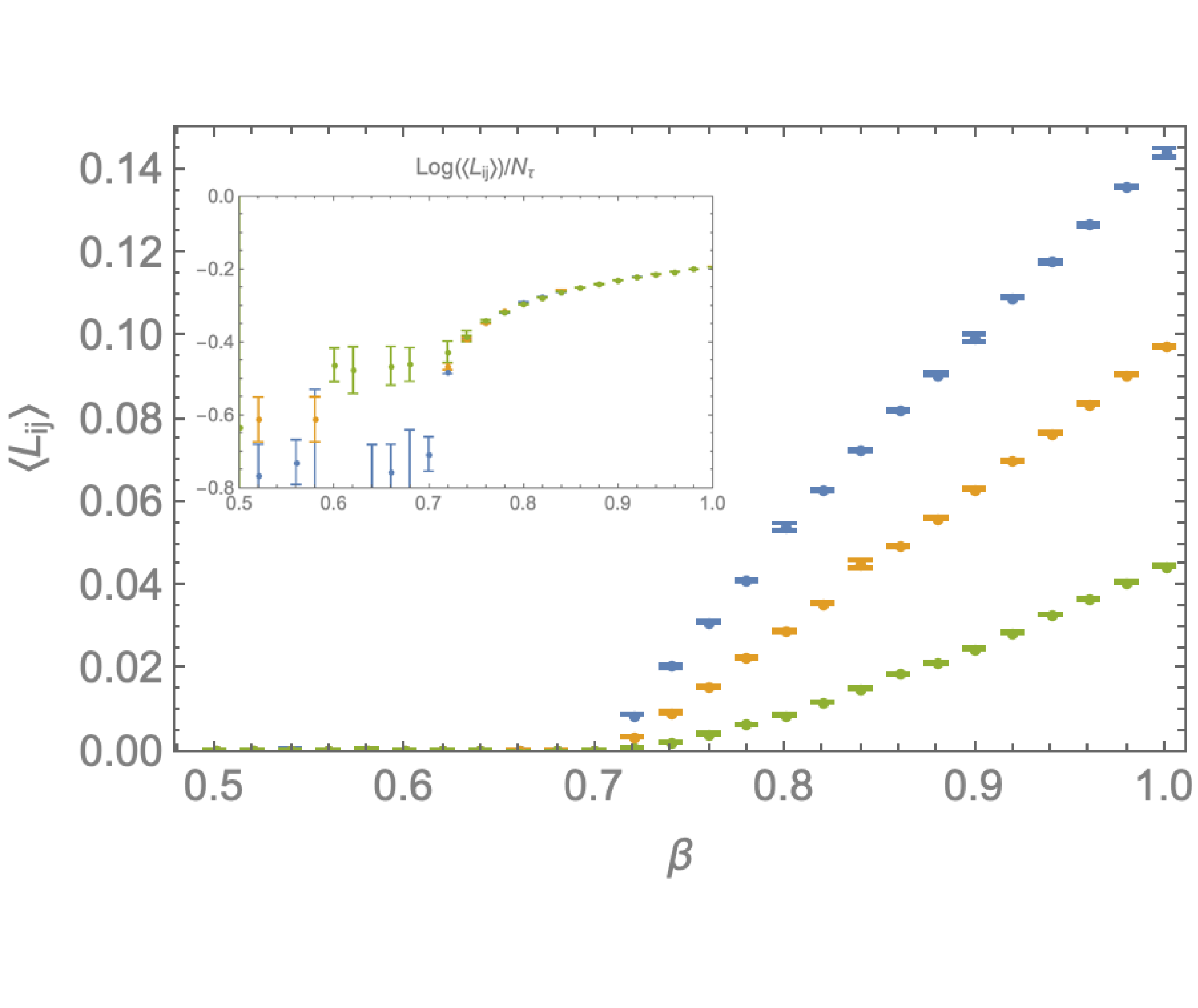}
         \caption{Line 5}
     \end{subfigure}
     \hfill
     \begin{subfigure}[b]{0.45\textwidth}
         \centering
         \includegraphics[width=\textwidth]{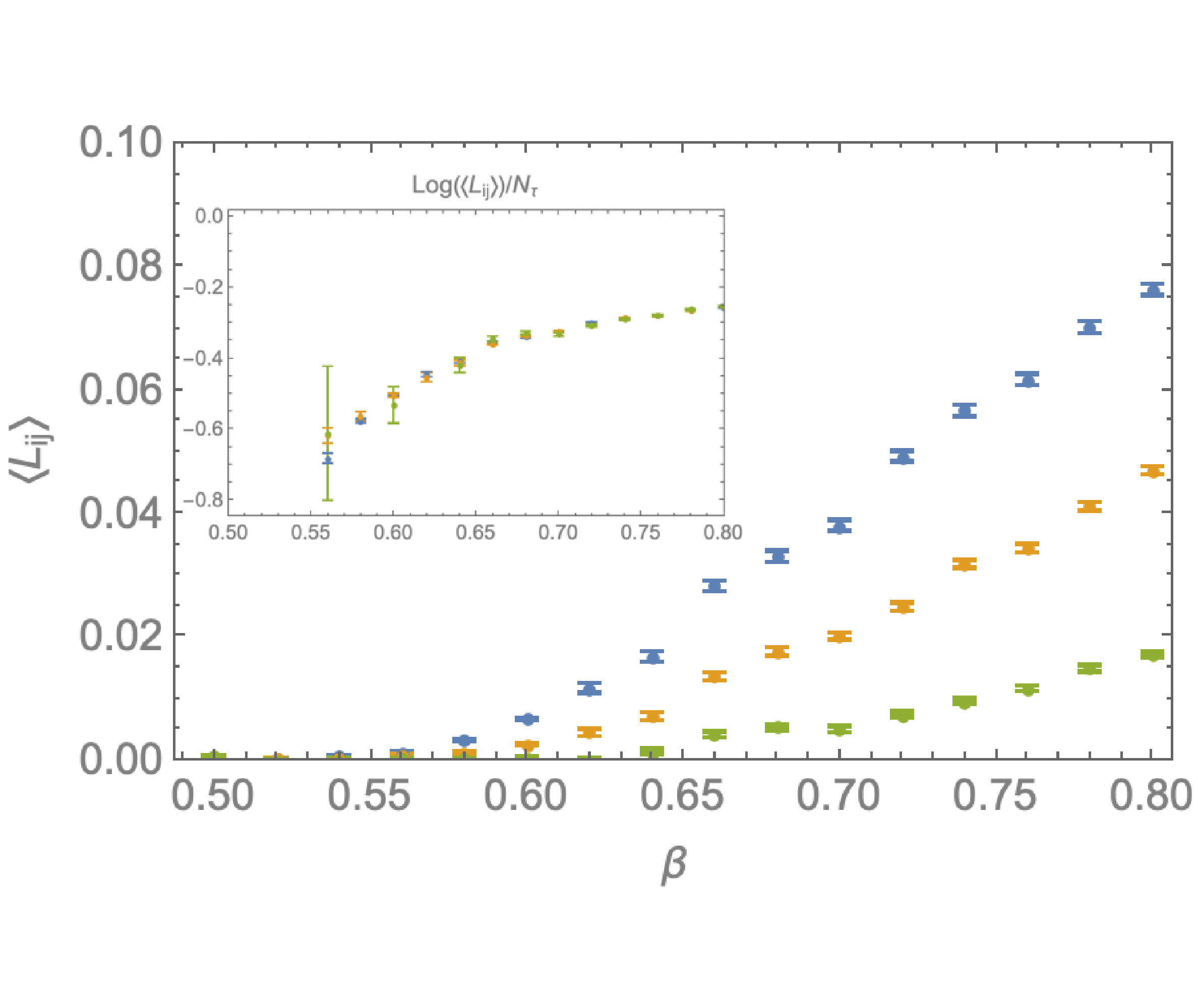}
         \caption{Line 6}
     \end{subfigure}
     \caption{Cuts represented in the time tubes phase diagram for charge $q=1$ above. The inset in each image is the quotient of the time tubes logarithm with the system temporal span $N_\tau$. This quantity is expected to be system size invariant. For strong coupling(small $\beta$) this sometimes fails (Lines $1,4,5$). The expected value for the time tubes can be so small in this regime that it is unfeasible to obtain enough data for the necessary precision. For weak coupling(large $\beta$) it is always observed. }
     \label{fig:TTcharge1_cuts}
\end{figure}

We begin our exploration of the Higgsed A tensor lattice gauge theory by surveying the full phase diagrams of the models specified by lattice action Eqs.~\ref{eq:gaugelatticeaction} and~\ref{eq:higgslatticeaction} with charge $q=1$ and $2$ Higgs fields. In particular, we present results from Monte Carlo simulations as a function of $\beta$ and $\kappa$ computing the analog of the (gauge invariant) Polyakov loop operator

\beq
\langle L_{ij} \rangle = \langle \prod_{n=0}^{N_t - 1} L_{ij\tau}(\vec{r}+n\vec{e}_\tau) \rangle. 
\eeq
Fig.~\ref{fig:TTcharge1} shows the variation of $\langle L_{ij} \rangle$ obtained from simulations on a $16^4$ hypercubic lattice for $q=1$. Broadly, $\langle L_{ij} \rangle$ is strongly suppressed for small $\beta$ and finite for larger $\beta$. One finds that $\log \langle L_{ij} \rangle/N_t$ collapses results for different system sizes. This quantity is plotted along cuts (indicated as dashed lines in Fig.~\ref{fig:TTcharge1}) in Fig.~\ref{fig:TTcharge1_cuts}. Here one sees that there is an abrupt step indicative of a first order transition along cuts $2-5$. The height of the step decreases as $\beta$ decreases and no such step is visible along cut $6$ which, instead, is smooth as a function of the couplings. Therefore the existence of a critical point can be surmised. The white solid line in Fig.~\ref{fig:TTcharge1} marks the extent of the first order line and the approximate location of the putative critical endpoint.

\onecolumngrid

\twocolumngrid

\begin{figure}[H]
    \centering
    \includegraphics[width=0.9\columnwidth]{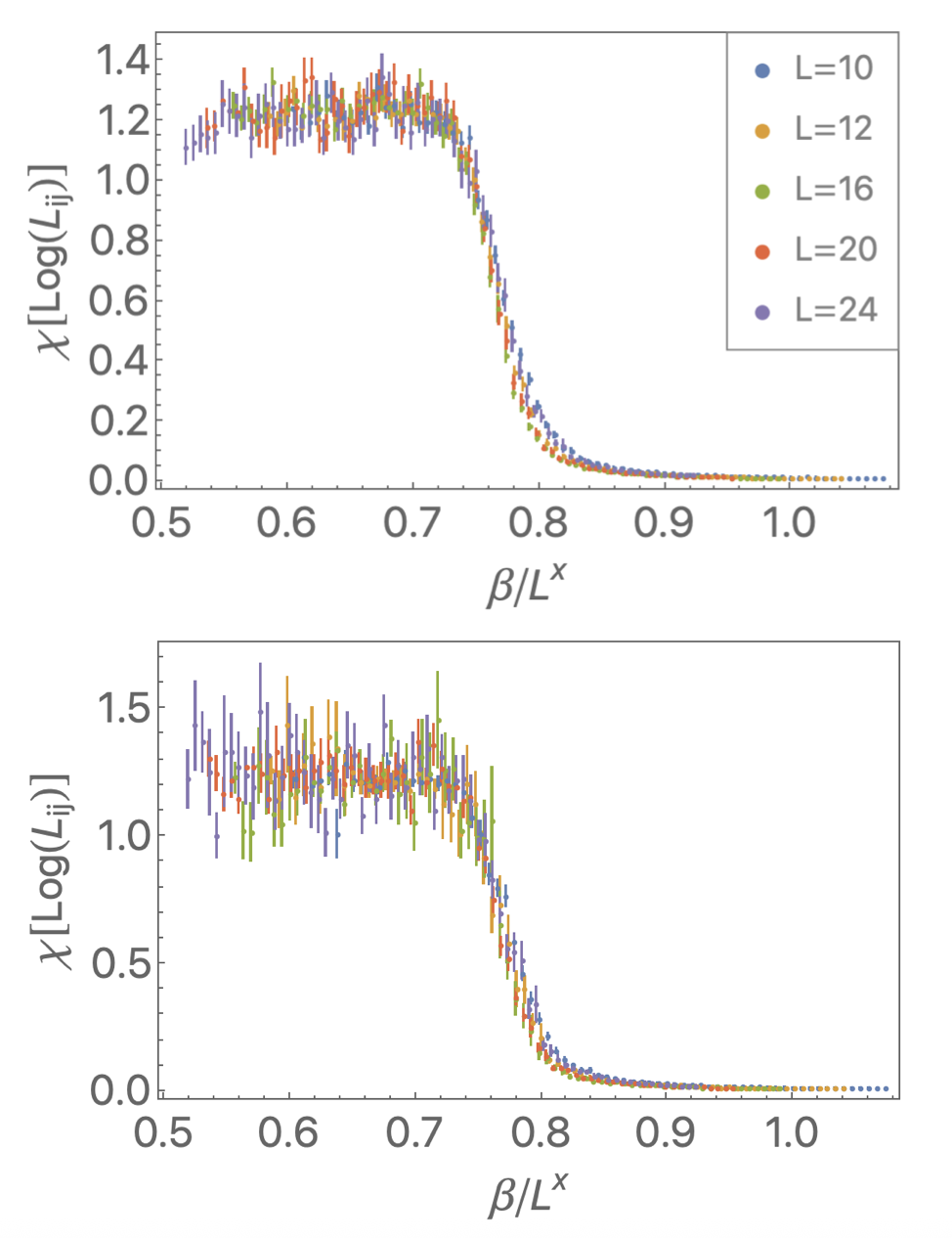}
    \caption{Susceptibility of the logarithm of the time tubes for $\kappa=0$ (top) and $\kappa=0.5$ (bottom). For both cases $x \approx 0.173$. A crossover between different regimes can be observed, but it does not survive the thermodynamic limit.}
    \label{fig:TT_chi_log}
\end{figure}

A cursory examination of Fig.~\ref{fig:TTcharge1} raises the possibility of a further transition, visible in $\langle L_{ij} \rangle$, for small fixed $\kappa$ as $\beta$ is varied. Looking instead at the horizontal cut $1$ in Fig.~\ref{fig:TTcharge1_cuts} no sharp feature is visible. This requires further investigation including a careful finite size scaling analysis. We considered various observables finding no clear signs that a $\beta$ driven transition is present in the thermodynamic limit. In contrast, we find a compelling case for a crossover. For example,  Fig.~\ref{fig:TT_chi_log} shows data for the susceptibility of the logarithm of $L_{ij}$ plotted against $\beta/L^x$ for both zero matter coupling, $\kappa$ and small finite $\kappa=0.5$ (the latter corresponding to cut $1$ in Fig.~\ref{fig:TTcharge1_cuts}). We find a collapse of data for different system sizes for exponent $x=0.173$. This result implies that the peak in this quantity drifts to larger $\beta$ as the system size increases so the naive weak coupling theory does not survive in the thermodynamic limit.  

\hspace{-2.5cm}

The conclusion from this study is that there is a single thermodynamic phase in the $q=1$ case but that there is a first order phase transition line similar to a liquid-gas transition line with a critical endpoint. This feature is also present in both the Abelian Higgs model in four dimensions (albeit accompanied by a deconfined phase for small $\kappa$, large $\beta$) and also in non-Abelian Higgs models (where, in common with the model considered here, there is a single phase). It would be of interest to investigate further the nature of the criticality at the endpoint but this is beyond the scope of the current investigation.

\begin{figure}[h]
    \centering
    \includegraphics[width=0.8\columnwidth]{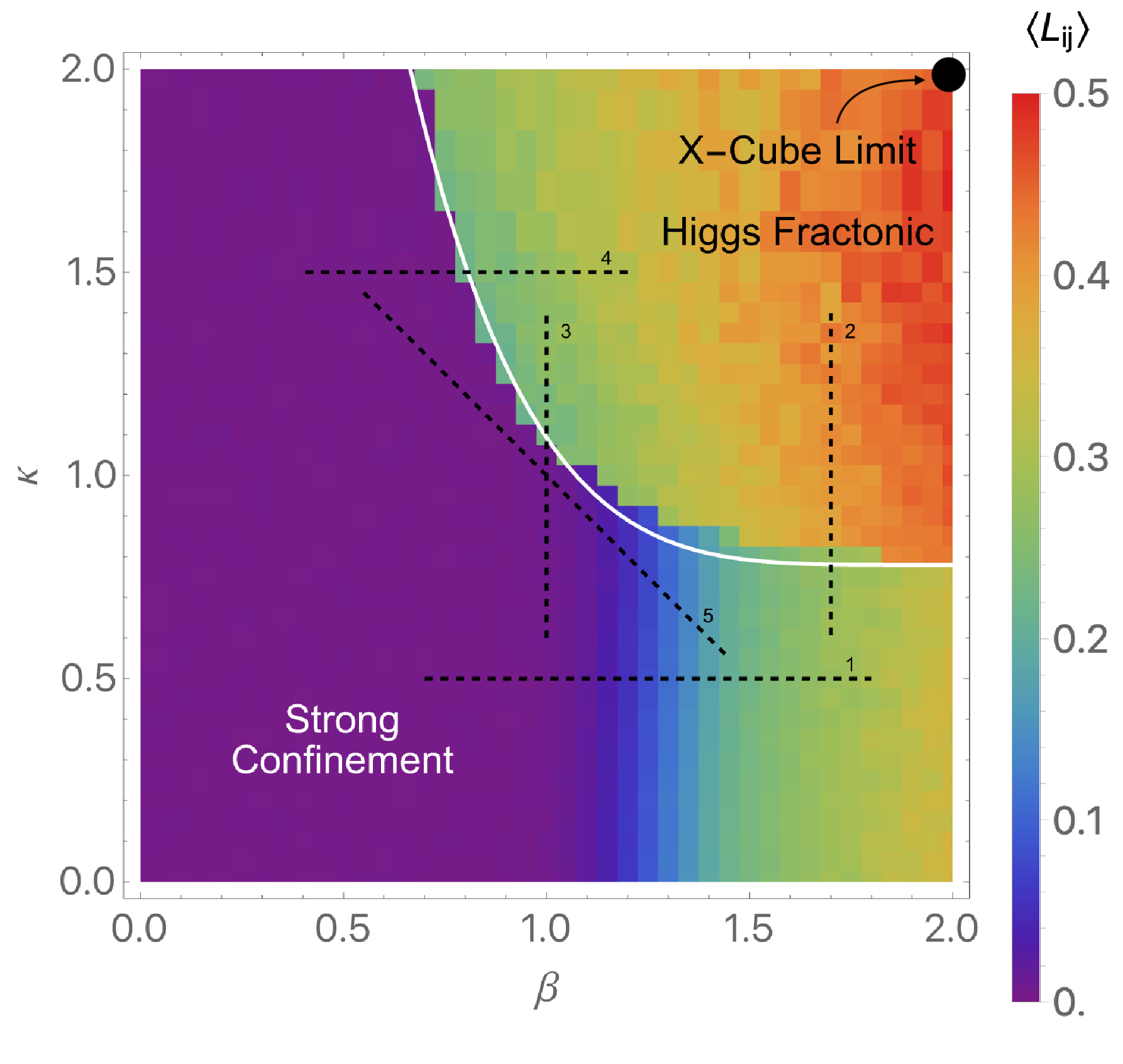}
    \caption{Phase diagram of the $4D$ charge $2$ higgsed hollow rank 2 gauge theory parametrized by $\beta$ and $\kappa$.The order parameter plotted is the time tube $\langle L_{ij}\rangle$. Unlike the charge $1$ case, there is a phase boundary extending to infinite $\kappa$ separating the Higgs phase from the fractonic confined phase.This phase diagram was obtained from Monte Carlo simulations on a $16^4$ lattice.}
    \label{fig:TTcharge2}
\end{figure}

\begin{figure}[H]
     \centering
     \begin{subfigure}[b]{0.45\textwidth}
         \centering
         \includegraphics[width=\textwidth]{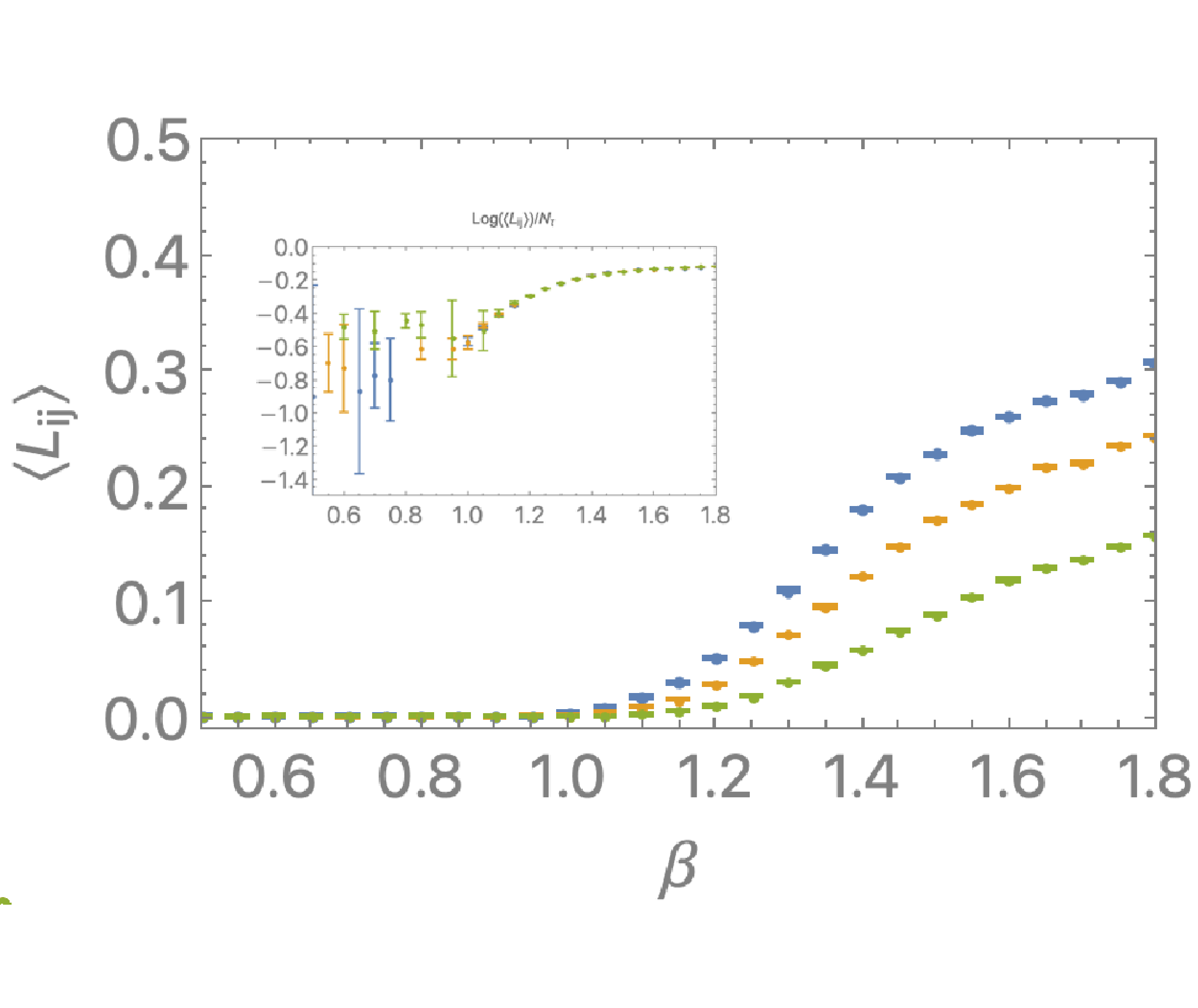}
         \caption{Line 1}
     \end{subfigure}
     \hfill
     \begin{subfigure}[b]{0.45\textwidth}
         \centering
         \includegraphics[width=\textwidth]{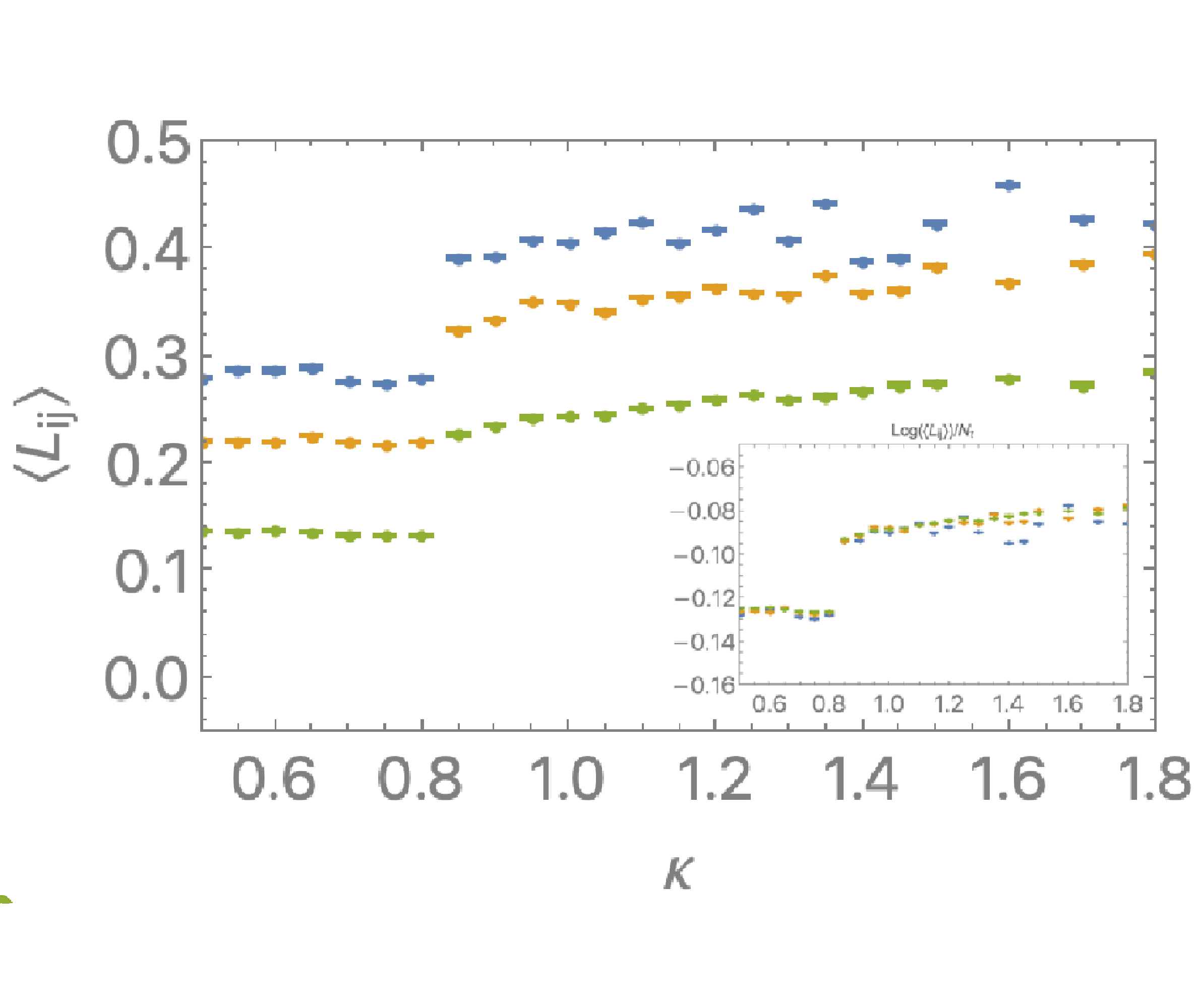}
         \caption{Line 2}
     \end{subfigure}
     \hfill
     \begin{subfigure}[b]{0.45\textwidth}
         \centering
         \includegraphics[width=\textwidth]{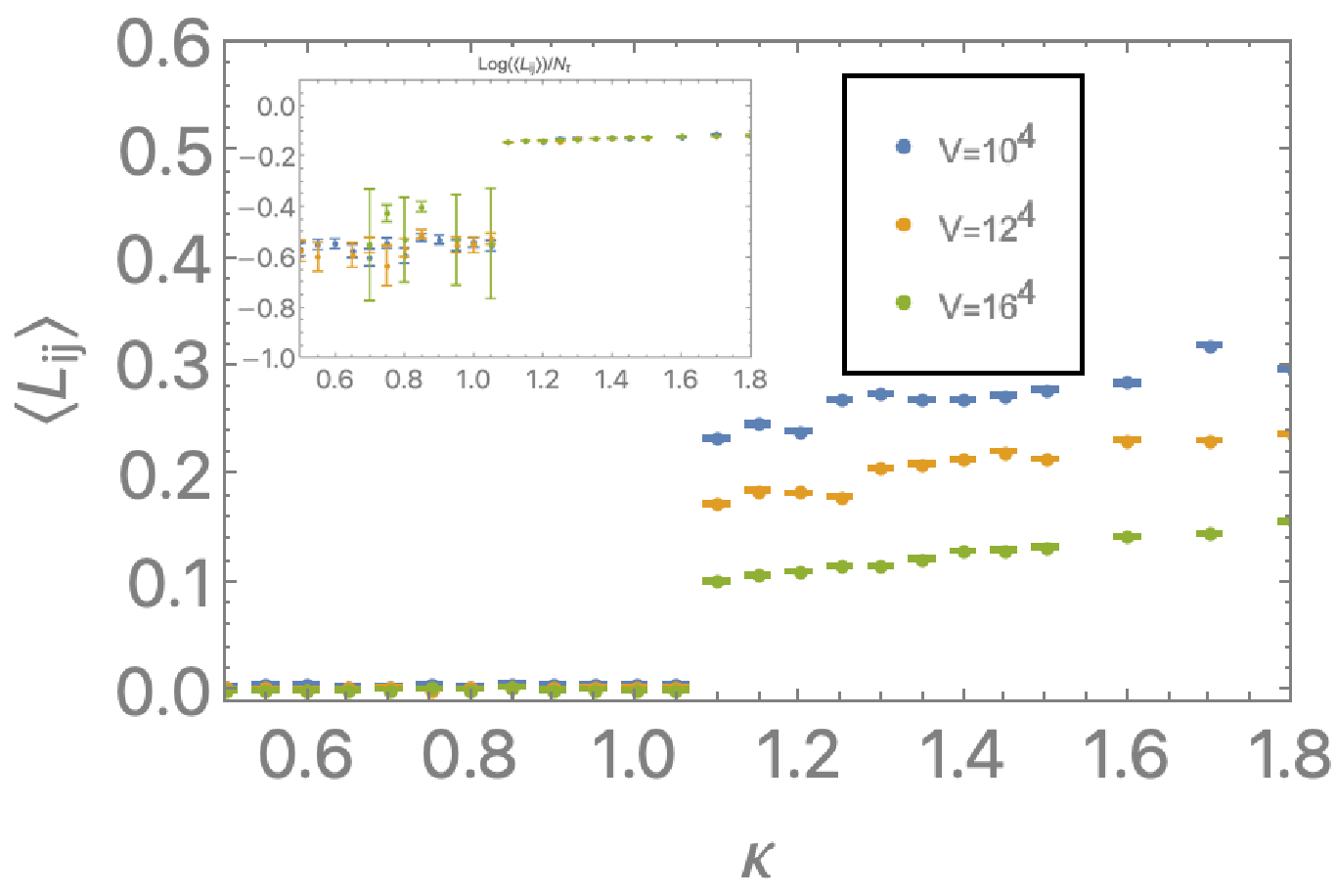}
         \caption{Line 3}
     \end{subfigure}
     \hfill
     \begin{subfigure}[b]{0.45\textwidth}
         \centering
         \includegraphics[width=\textwidth]{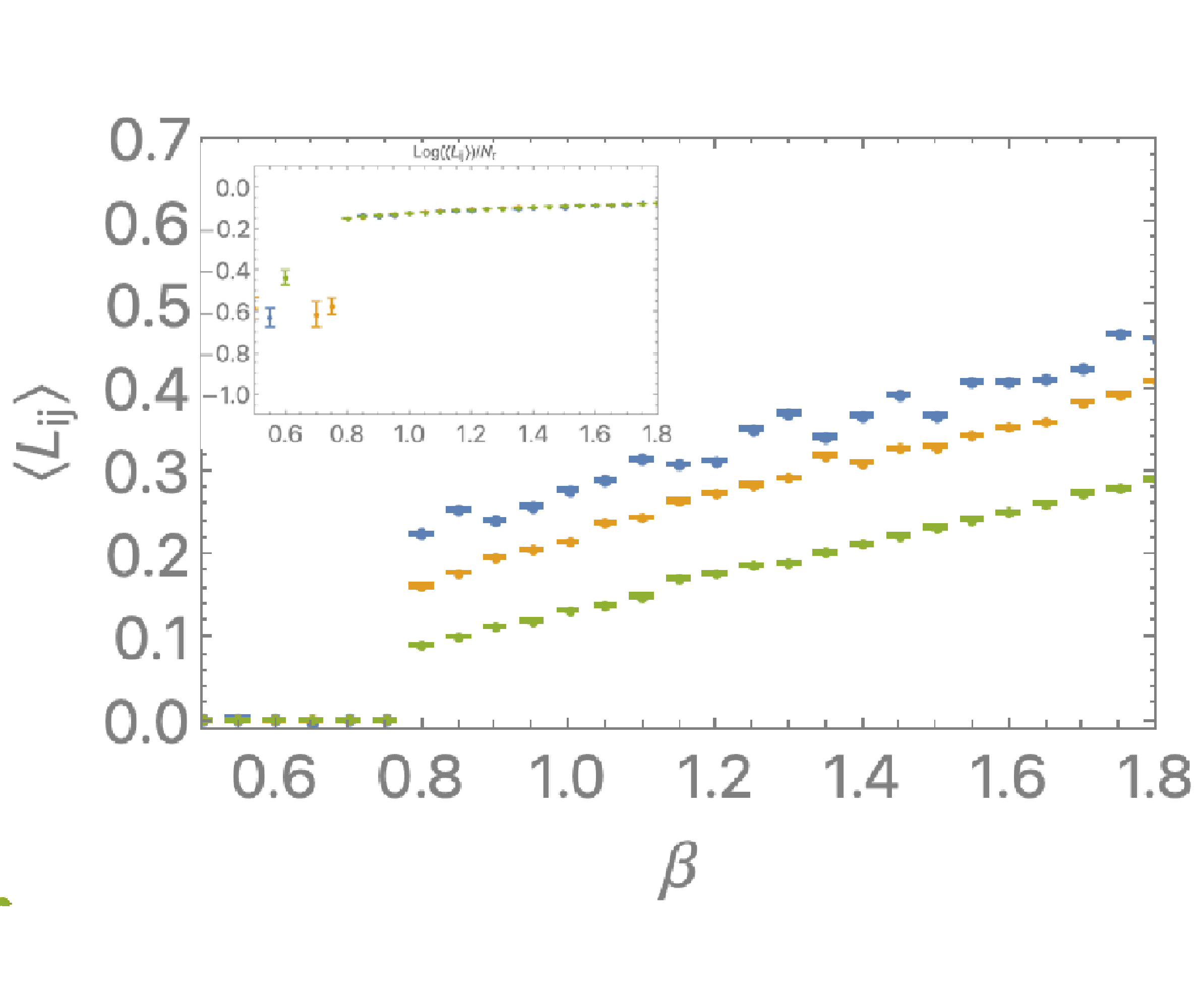}
         \caption{Line 4}
     \end{subfigure}
     \hfill
     \begin{subfigure}[b]{0.45\textwidth}
         \centering
         \includegraphics[width=\textwidth]{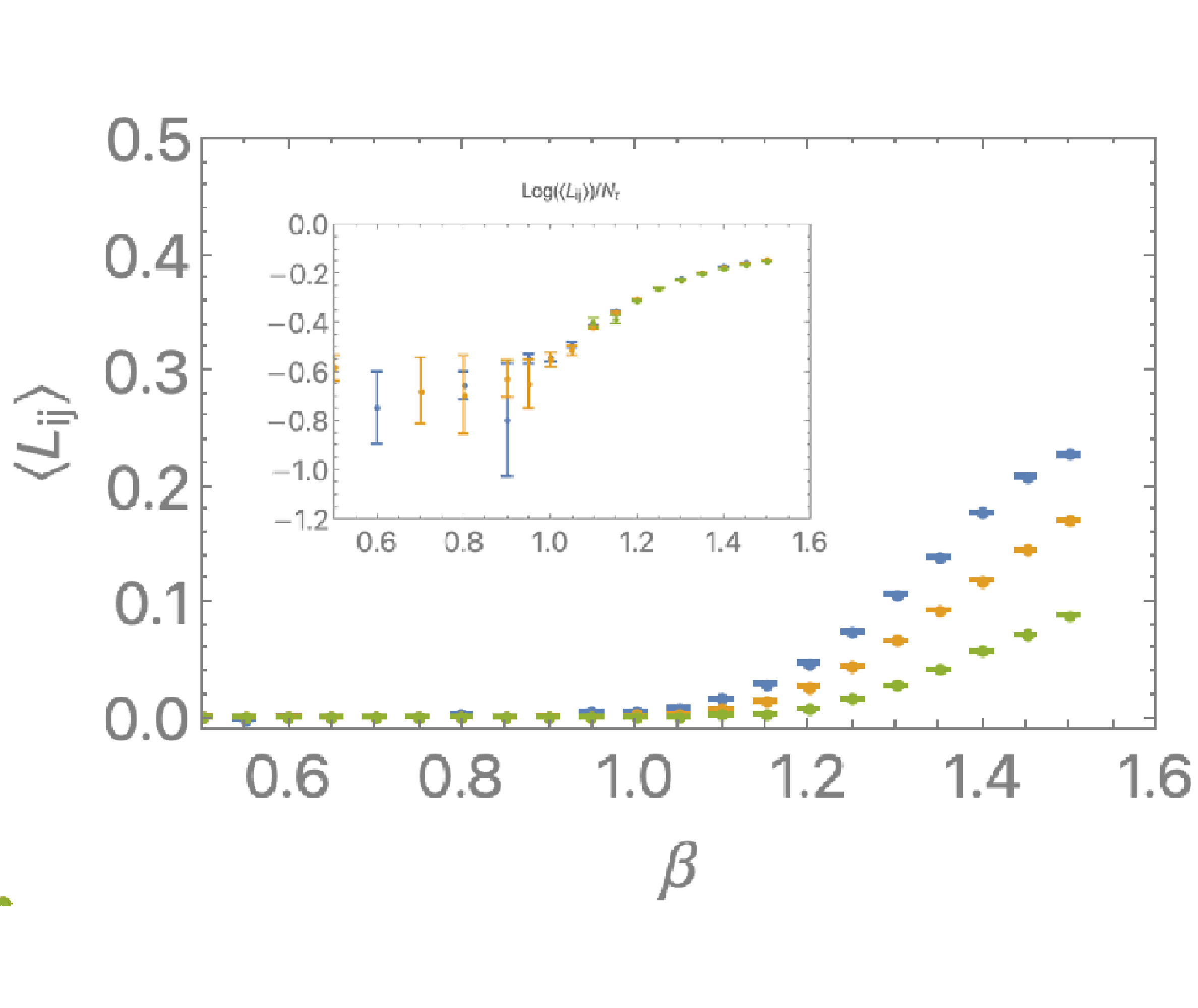}
         \caption{Line 5}
     \end{subfigure}
     \caption{Cuts represented in the time tubes phase diagram for charge $q=2$ above. The inset in each image is the quotient of the time tubes logarithm with the system temporal span $N_\tau$. This quantity is expected to be system size invariant. For strong coupling(small $\beta$) this sometimes fails for the sames reason as before, but is always observed for weak coupling(large $\beta$)}
     \label{fig:TTcharge2_cuts}
\end{figure}

Now we turn to the charge $q=2$ case, where a similar study was employed. Fig.~\ref{fig:TTcharge2} shows the variation of $\langle L_{ij} \rangle$ obtained from simulations on a $16^4$ hypercubic lattice for $q=2$. We again can see that $\langle L_{ij} \rangle$ is strongly suppressed for small $\beta$ and finite for larger $\beta$. The main difference here is the first order transition extends until infinite $\kappa$ and lack of critical point. This transition can be clearly seen in the cuts 2-4 (indicated as dashed lines in Fig.~\ref{fig:TTcharge2}) which are plotted in Fig.~\ref{fig:TTcharge2_cuts}. Cuts 1 and 5 do not cross this transition and have a smooth crossover, similar to the $q=1$ case. We have already discussed the infinite matter coupling limit in section \ref{subsection:InfMatter} and concluded it should have a phase transition. We can then surmise that the observed first order transition extends until infinity. The white solid line in Fig.~\ref{fig:TTcharge2} serves as a guide for the eye for the first order transition location. The expected location of the X-Cube limit is also marked.

\onecolumngrid

\twocolumngrid

\par It can be concluded then that for $q=2$ there is now an additional thermodynamic phase: a distinct Higgs phase analogous to that of the Abelian Higgs model in four dimensions, where a similar behavior is observed for charge $q=2$ as well. However, the rank-$2$ Higgs phase has fractonic topological order.

\par As a first attempt to characterize this newfound fractonic Higgs phase we study the extended time cubes introduced in the previous section at different points in the phase diagram.

\begin{figure}[H]
\begin{center}
  \includegraphics[width=\columnwidth]{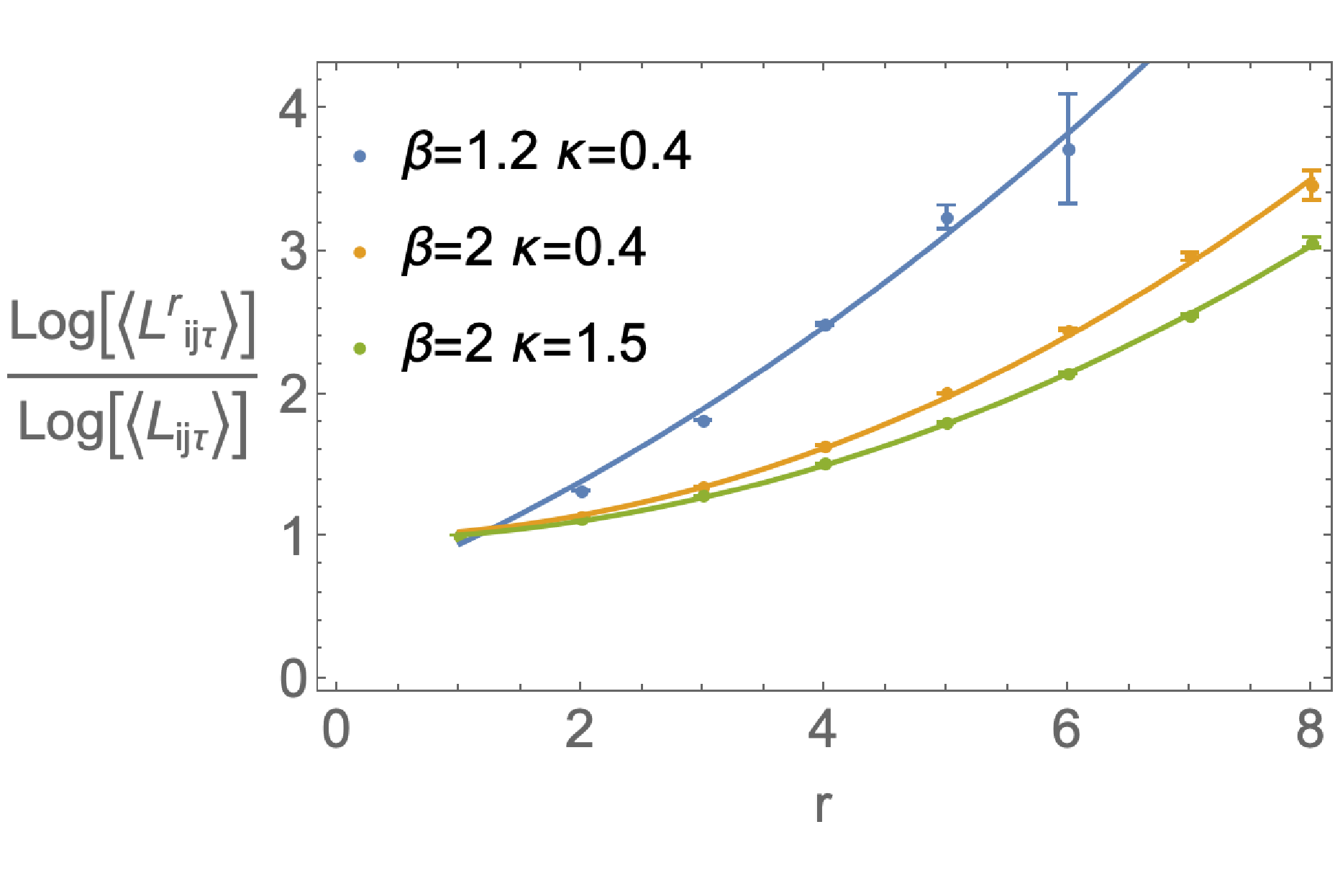}
  \end{center}
  \caption{Normalized logarithm of the extended time cubes for a system size V=24x24x24 in different locations of the phase diagram for charge $q=1$. We have fitted the data to quadratic order and plotted the result. The fitting results were $0.034r^2+0.34r+0.57$, $0.039r^2+0.046r+0.99$ and $0.032r^2+0.005r+0.97$ for the blue, orange and green curves respectively. }
\end{figure}

\par As can be seen from the previous figure, there were no qualitatively different results for the extended time-cubes in the fractonic Higgs phase, being well described by a quadratic function. This motivates the study of correlation functions as a way to characterize the different phases, which we do in the following section. 

 \vspace{-0.5cm}
\subsection{Correlation functions and pinch points}
 \vspace{-0.5cm}

\par So far we have found evidence for a single phase in the $q=1$ hollow rank-two theory coupled to bosonic matter and for two phases in the $q=2$ case. In this section, we consider the problem of characterizing these phases through correlation functions. 

\par In particular, we compute the higher-rank electric tensor correlation function 
\be
 C_{ij}^{kl}\left(r \right) = \langle L_{ij\tau}\left(r \right)L_{kl\tau}^{\dag}\left(0 \right) \rangle  - \langle L_{ij\tau} \rangle \langle L_{kl\tau}^{\dag} \rangle.
\label{eq:cijijsce}
\ee

\par
\begin{figure}[H]
    \centering
    \includegraphics[width=1\columnwidth]{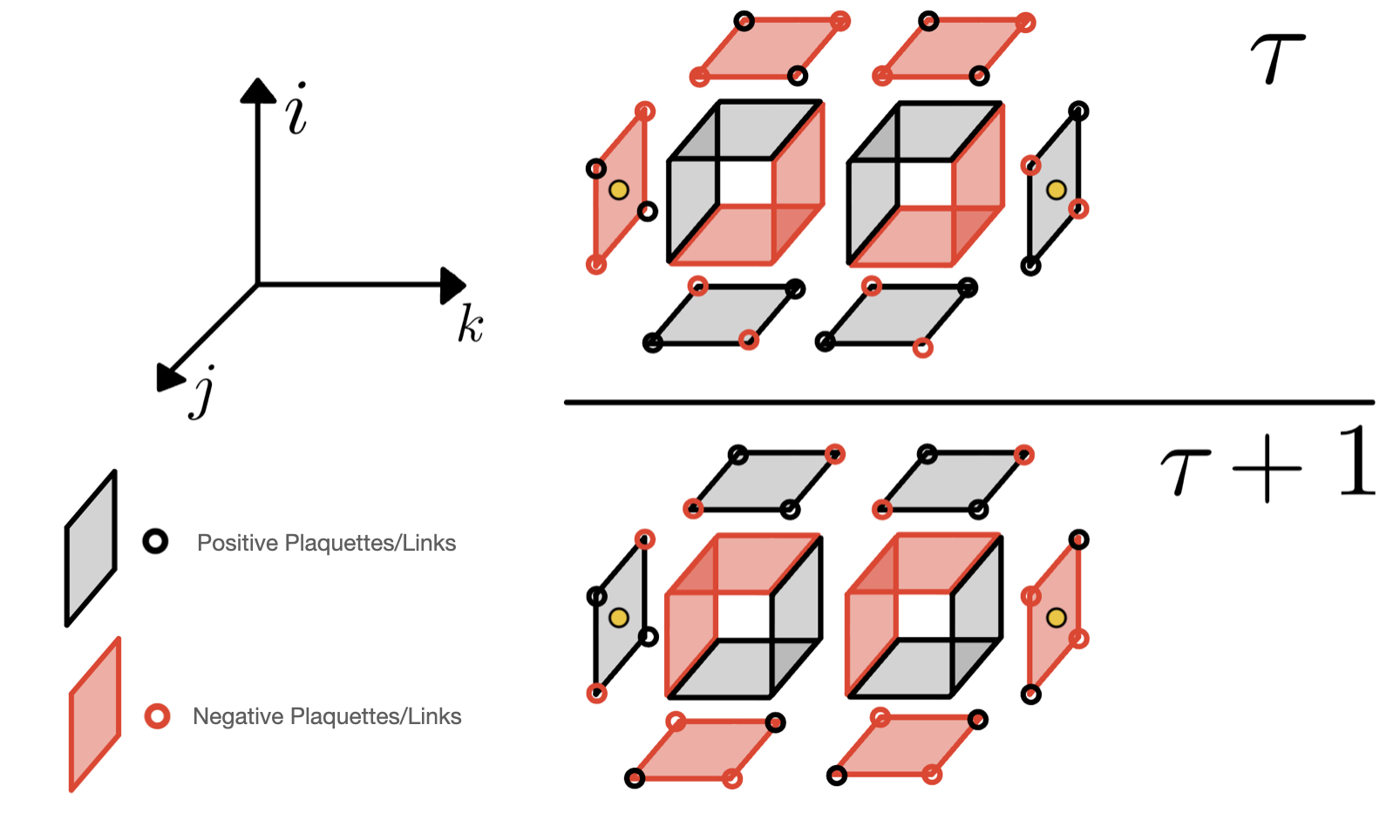}
    \caption{Strong coupling higher-order gauge variable structure for the correlation function $C_{ij}^{ij}$ when the electric tensor are separated in the $k$ direction. In the figure we represent the time links as dots on the corners of the space plaquettes. Since the electric field tensors are temporal-spatial objects, we present the two different time slices in which they live. The electric tensors in question are represented with a yellow dot. }
    \label{fig:HighOrderCorr}
\end{figure}

 Analytically, these functions can be studied in the strong coupling regime similarly to the time tubes discussed earlier. To first-order in the strong coupling expansion there is no separation dependence, making it necessary to include higher order terms. This is done by finding the smallest possible gauge invariant combination that cancels every gauge variable and includes both the electric tensor we want to study. One example of such a construction can be found in Fig.~\ref{fig:HighOrderCorr}.

\par There is a $\beta/2$ factor for each individual time cube $L_{ij\tau}$ and space cube $L_{[ij]k}$ that is contained in the large structure that joins the electric tensors. For the case presented in Fig.~\ref{fig:HighOrderCorr} it is simple to see that the correlation function at distance $r$ to leading order in the strong coupling expansion is

\be
C_{ij}^{ij} \left( r \vec{e_k} \right) = 2^{|r|} \left( \frac{\beta}{2} \right)^{4|r|} .
\ee

\par There are different possible combinations of $L_{ij\tau}$ and $L_{[ij]k}$ that lead to the same results. Those are taken into account with the factor $2^{|r|}$. Results for other correlation functions of interest are summarized in the table below:

\begin{table}[H]
\centering
\begin{tabular}{|l|l|}
\hline
$C_{ij}^{ij} \left( r \vec{e_i} / \vec{e_j} \right)$ &  $2^{|r|-1}\left( r-1 \right) \left( \frac{\beta}{2} \right)^{4(|r|+2)}$ \\ \hline
$C_{ik}^{kj} \left( r \vec{e_k} \right)$ & $2^{|r|-1}\left( r-1 \right) \left( \frac{\beta}{2} \right)^{4(|r|+2)}$ \\ \hline
$C_{ik}^{kj} \left( r \vec{e_i} / \vec{e_j} \right)$ & $2^{|r|} \left( \frac{\beta}{2} \right)^{4(|r|+1)}$  \\ \hline
\end{tabular}
\end{table}

\par
\begin{figure}[H]
    \centering
    \includegraphics[width=0.8\columnwidth]{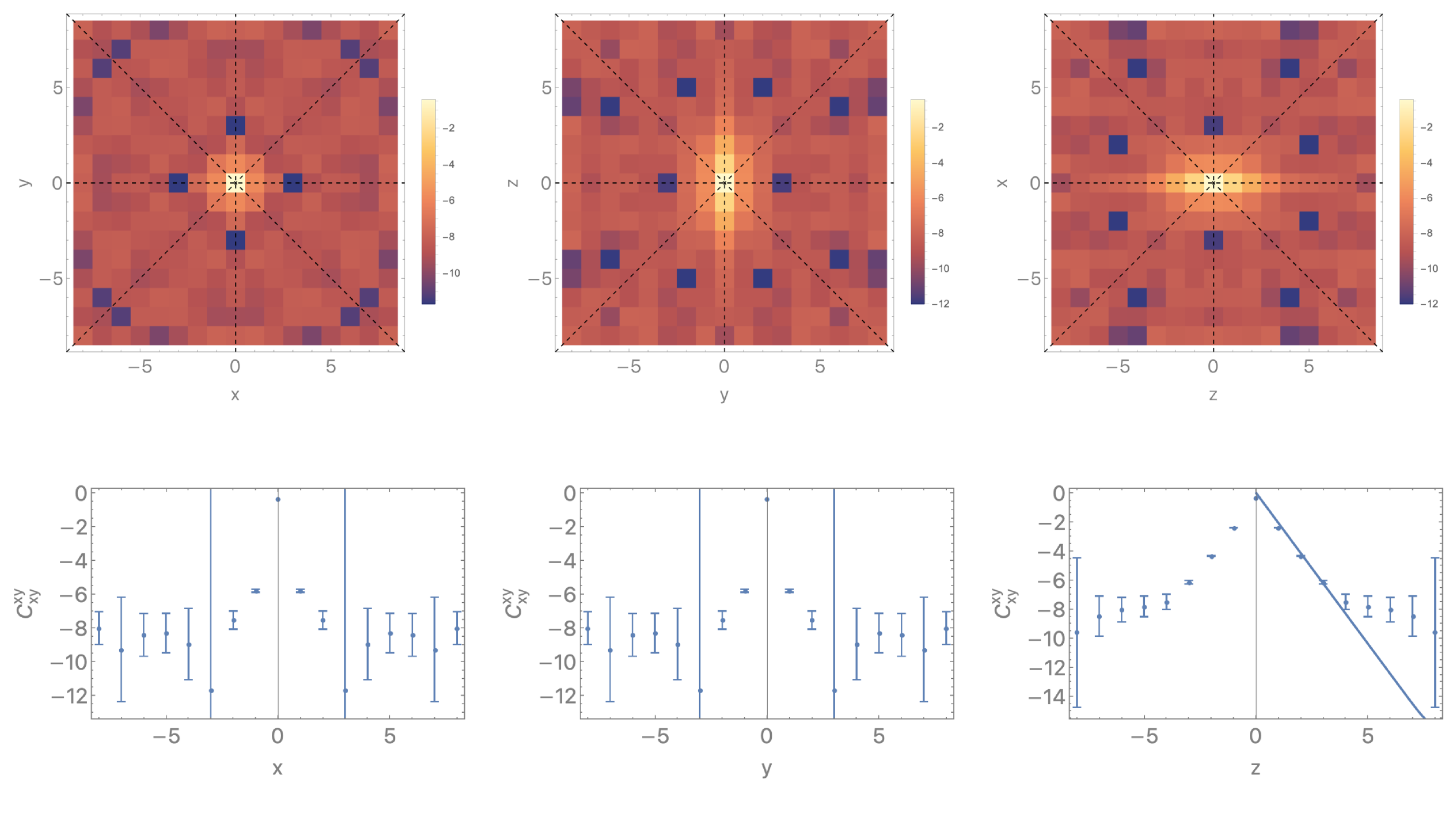}
    \caption{Logarithm of the correlation functions $C_{xy}^{xy}$ in real space for charge $q=2$ and couplings $\beta=1$ and $\kappa=0$. The top plots are along the $z=0$, $x=0$ and $y=0$ planes. The bottom cuts are along the $x$, $y$ and $z$ axis. For the latter we compare the numerical results with the strong coupling result and find agreement between them. For the former two this comparison is not made because the correlations decay too quickly (which is expected from the strong coupling expansion).  }
    \label{fig:realspacecorr}
\end{figure}

 These correlation function were also obtained numerically using the same Monte-Carlo algorithm used throughout this work. Results are shown in Fig.~\ref{fig:realspacecorr} for coupling $\beta=1$ and $\kappa=0$ and for $C_{xy}^{xy}$. Here we chose an intermediate coupling as the absolute magnitude of the correlation function is smaller for stronger couplings such that it harder to distinguish from noise for $\beta$ much below $1$. We observe a four-fold symmetry in the $xy$ plane and two-fold symmetry in the $yz$ and $xz$ planes. The decay along the $z$ direction is slow enough that we may fit the result to the strong coupling prediction (Eq.~\ref{eq:cijijsce}) finding good agreement for small distances. 
\onecolumngrid

\twocolumngrid

\par Similar findings were found for the $C_{xy}^{xz}$ correlator including agreement with the strong coupling expansion. 

\par
\begin{figure}[H]
    \centering
    \includegraphics[width=0.8\columnwidth]{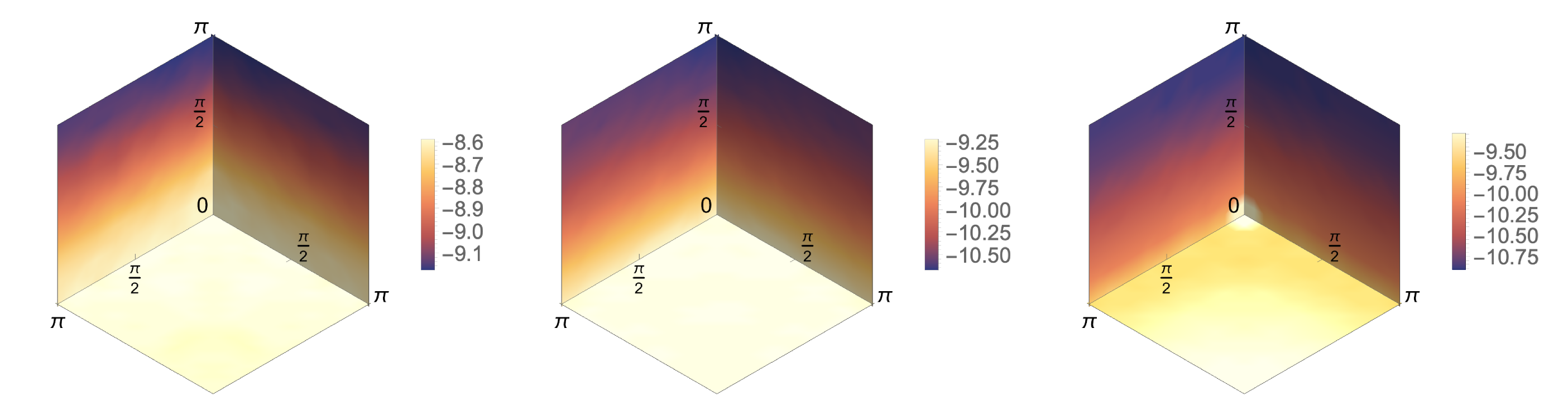}
    \caption{Logarithm of the correlation functions $C_{xy}^{xy}$ in momentum space for charge $q=2$ for different points in the phase diagram. In the left we have $\beta=1$,$\kappa=0$, in the middle $\beta=2$, $\kappa=0.5$ and on the right  $\beta=2$, $\kappa=1.5$. }
    \label{fig:corr_fourier_3d}
\end{figure}

\begin{figure}[H]
    \centering
    \includegraphics[width=0.8\columnwidth]{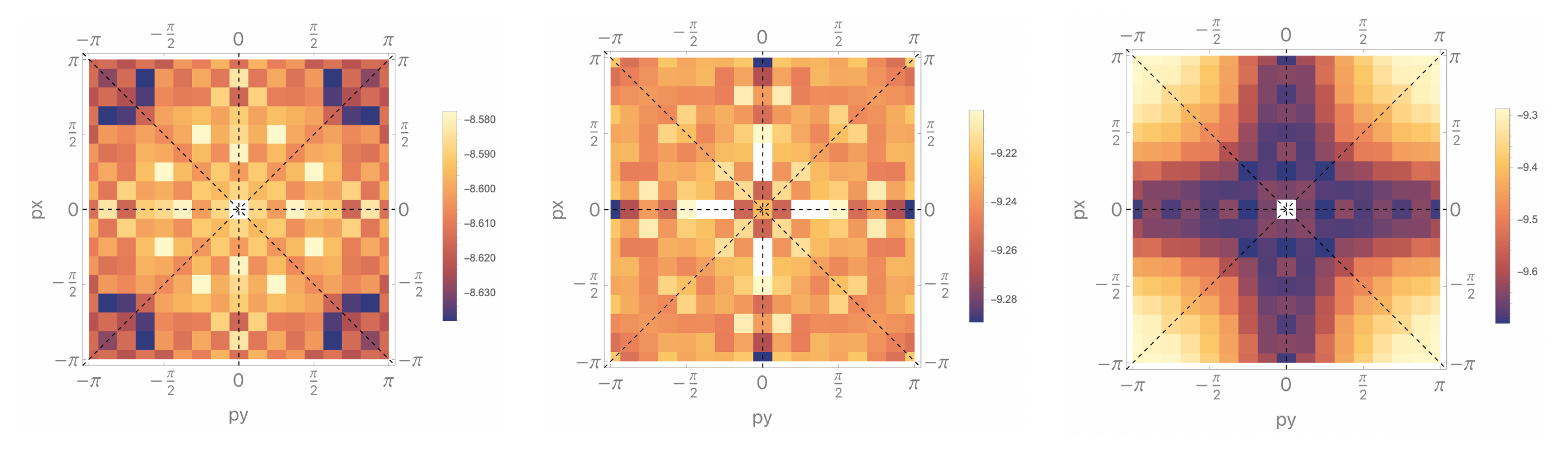}
    \caption{Same plots as the figure above, but now with focus on the plane $p_z = 0$.}
    \label{fig:corr_fourier_2d}
\end{figure}

 Figs.~\ref{fig:corr_fourier_3d} and \ref{fig:corr_fourier_2d} show the $C_{xy}^{xy}$ correlation function in momentum space for the $q=2$ model at three different points in the phase diagram. The two plots on the left are for the confining phase respectively at intermediate and weak coupling. The plot on the right in each figure is taken within the Higgs phase. We have discussed the strong coupling result in real space. In momentum space the intensity is concentrated around the $p_x-p_y$ plane with a four-fold symmetry but with few features in that plane. 

\par The case of weak coupling deserves further discussion. The naive continuum limit of non-hollow rank two theory reveals electric tensor correlation functions with pinch point singularities \cite{prem2018} with four-fold symmetry. This is reminiscent of the $U(1)$ spin liquid coming from the deconfined phase of a pure $U(1)$ gauge theory which also has pinch-points but with with two-fold symmetry. Pinch point provide a sharp experimental signature that could be used to diagnose tensor Coulomb spin liquids in, for example, neutron scattering experiments. We have shown in Appendix~\ref{sec:weak_coupling_pinch_points} that similar features are present in the naive weak coupling limit on the lattice in the theory under consideration here. 

We have also presented evidence that instanton proliferation destroys the naive weak coupling regime. However, this evidence from the Monte Carlo required careful finite size scaling. Inspection of the correlations at small $\kappa$ and $\beta=2$ that would exhibit pinch points in the naive weak coupling regime instead show weak signs of a ridge along the $p_x$ and $p_y$ axes which is suggestive of the pinch point albeit only as a very weak feature visible here only on a logarithmic scale. It is expected that this feature scales away in the thermodynamic limit. 

\onecolumngrid

\twocolumngrid

The results presented above are expected to carry over to the phase of the $q=1$ model. We now turn our attention to the other distinct phase: the Higgs phase in the $q=2$ theory. Here the correlations again are concentrated in a plane but now the correlations in that plane show a distinctive ``cross" feature along the $p_x$ and $p_y$ axes that is not present in the confined phase (see Fig.~\ref{fig:corr_fourier_2d} $-$ right panel).

\section{Discussion}

Inspired by the continuum field theory counterpart that has been extensively studied over the last few years, in this paper we have carried out a systematic study of the hollow rank-two gauge theory in four dimensions on a lattice. 

We have discussed the appropriate lattice Hamiltonian and action from first principles introducing generalizations of the plaquette variables one sees in one-form lattice gauge theory. We have also introduced scalar matter into the theory carrying charge $q=1,2$. 

We have analysed various limiting cases of the hollow rank two Higgs theory. One interesting case is the limit of weak gauge coupling $\beta\rightarrow \infty$ controlled by matter-gauge coupling $\kappa$. We saw that the resulting statistical mechanics model has $U(1)$ subsystem symmetries and, via Monte Carlo simulation, we observed that there is a first order transition at some $\kappa^*$ separating the paramagnet from a long-range ordered phase that spontaneously breaks subsystem symmetry. 

We also studied the case where the matter coupling goes to zero. This pure gauge theory has a strong confinement regime at strong coupling that can studied using perturbation theory where pairs of dipoles are confined by a linear potential. The usual continuum field theory for which we study the lattice analog has a number of peculiar properties that we describe in an appendix including pinch point correlations. However, this weak coupling picture is destroyed by instantons through a higher rank version of the Polyakov mechanism. To our knowledge, this work is the first to show, through a numerical solution of the Debye-H\"{u}ckel equations, that strong confinement survives to weak coupling in this theory via a derivation of effective action for the instantons. We further bolster this result by studying the electric field correlations that show pinch points in the continuum weak coupling gauge theory. As expected no pinch points are present.

These findings for the limiting cases of the $q=1$ theory reveal a single phase transition. Through Monte Carlo simulations of the full theory we find that the transition extends to finite $\beta,\kappa$ terminating at a critical endpoint that would be interesting to explore further. In other words, this theory has a single phase that we characterized numerically and in the zero matter coupling setting. 

The $q=2$ theory has two phases. This can be seen already by taking the $\kappa\rightarrow \infty$ limit where, similarly to the charge two $U(1)$ gauge theory in four dimensions, there is a residual global $\mathbb{Z}_2$ degree of freedom. In this limit there is a $\beta$ driven phase transition and Monte Carlo simulations of the full theory show that this phase boundary meets the phase boundary originating from the infinite $\beta$ limit. This large-$\beta,\kappa$ Higgs phase is the dynamical gauge theory extension of the X-cube model.  

There are various potentially interesting open problems that are beyond the scope of this study. It would be interesting to explore the full rank-2 scalar charge theory with diagonal terms as the mechanism of partial confinement that leads to the hollow theory studied here \cite{Ma_2018}. Also it would be interesting to see, via numerical simulation, that the weak coupling continuum theory survive in higher dimensions. More importantly, can confinement be avoided in four dimensions coupling to fermionic matter? What is the behavior on a lattice of other types of rank-$N$ theory? Recently, some of us revealed the presence of boundary phase transitions in generic Abelian higher form and non-Abelian gauge theories coupled to scalar matter \cite{chung2024}.
What becomes of this transition in the case of higher rank gauge theories?

 \section*{Acknowledgements}
PR and PB acknowledge support  by FCT under contracts: LA/P/0095/2020, LaPMET, Laboratory of Physics for Materials and Emerging Technologies; and UID/04540, CeFEMA, Centre of Physics, Engineering and Advanced Materials.
PM acknowledges funding from the CNRS. MU is supported by KAKENHI Grants No. JP22H01147 and No. JP23K22418 from JSPS of Japan.

\appendix
 
 \section{Derivation of the lattice action}
 \label{sec:app_action}

In this appendix, we derive a lattice action for Hamiltonian $H=H_{\rm Gauge}+H_{\rm Higgs}$ given in the main text and repeated here for convenience.
\be
H_{\rm Gauge} = \sum_{\vec{r}, i<j} \frac{1}{g_e^2} E_{ij}^2+\frac{1}{g_m^2}( L_{[ij]k}(\vec{r})+L_{[ij]k}^\dag(\vec{r})).
\ee
and
\be
H_{\rm Higgs} = \sum_{\vec{r}} \frac{L(\vec{r})^2}{2M} - V \sum_{i<j} \cos(\Delta_i \Delta_j \theta (\vec{r}) + q A_{ij}(\vec{r}))
\label{eq:higgs2}
\ee
with Gauss law constraint
\be
G(\vec{r}) = \sum_{i< j} \Delta_i \Delta_j E_{ij}(\vec{r}) =  q L(\vec{r}). 
\ee
 Periodic boundary conditions are assumed, natural units will be used throughout and lattice constant $a$ is set to $1$. 
 
 Here we implement the Gauss law by restrict to the constrained subspace in the derivation of the action through projector 
\begin{align}
P &= \int D \phi \exp \left( i\sum_{\vec{r}} G(\vec{r}) \phi(\vec{r})\right)\nonumber\\ 
&= \int D\phi \exp \left( i\sum_{\vec{r}} \sum_{i<j} \Delta_i^b \Delta_j^b E_{ij}(\vec{r}) \phi(\vec{r})\right).
\end{align}

\par where $D\phi =\prod_{\vec{r}} \frac{d\phi(\vec{r})}{2\pi}$
\par The partition function, in the physical subspace, is then
\begin{align}
Z &= {\rm Tr }\left[ \exp(-\beta H)P\right] = {\rm Tr}\left[ (\exp\left(-\frac{\beta}{N}H\right)P )^N \right] \nonumber \\ 
&= \int DA \prod_{\tau=1}^N \langle A_\tau \vert \exp \left( -\Delta\tau H\right)P\vert A_{\tau+1}\rangle 
\end{align}
where a time slicing has been carried out with $ \Delta\tau = \beta/N$. 

We then split the Hamiltonian into two parts $H=H_A + H_E$, where $H_A$ contains the terms that depend on the gauge variables $A_{ij}$, $H_E$ depending on the conjugate variables $E_{ij}$. Therefore, for small $\Delta\tau$ we have 
\begin{align}
\exp \left( -\Delta \tau H\right)&=\exp \left( -\Delta \tau H_A-\Delta \tau H_E\right)\nonumber \\
&\simeq \exp \left( -\Delta \tau H_A\right)\exp \left( -\Delta \tau H_E\right).
\end{align}

Then we insert resolutions of the identity in the basis of $E$ eigenstates at each time slice
\begin{align}
Z \simeq \int DA \prod_{\tau=1}^N \langle A_\tau& \vert\exp \left( -\Delta\tau H_A\right)\exp \left( -\Delta\tau H_E\right)P \vert A_{\tau+1}\rangle \nonumber \\
=\int DA \prod_{\tau=1}^N \sum_{E_\tau} & \langle A_\tau \vert \exp \left( -\Delta\tau H_A\right) \vert E_\tau \rangle \nonumber\\
&\langle E_\tau \vert \exp \left( -\Delta\tau H_E\right)P \vert A_{\tau+1}\rangle \nonumber \\
= \int DA D\phi \sum_{E_{\tau}}&\prod_{\tau=1}^N \exp \left( -\Delta\tau H_{A_\tau}-\Delta\tau H_{E_\tau}\right. \nonumber \\ 
  &-i\sum_{\vec{r}} \sum_{i<k} \Delta_i^b \Delta_j^b E_{ij}^\tau\phi_\tau(\vec{r})  \nonumber \\
  &+i\sum_{i<j,\vec{r}}A_{ij}^\tau (\vec{r})E_{ij}^\tau (\vec{r}) -A_{ij}^{\tau+1} (\vec{r})E_{ij}^\tau (\vec{r})  )     \nonumber 
\end{align} 

Rearranging and setting $\theta_\tau(\vec{r})=-A_0^\tau (\vec{r})$ we find
\begin{align}
Z  = \text{const} \int DA  \sum_{E_{\tau}} \exp ( \sum_{\tau=1}^N \sum_{\vec{r} ,i<j }&  \frac{\Delta \tau}{g_m^2} (L_{[ij]k}^\tau (\vec{r})+L_{[ij]k}^{\tau \dag }(\vec{r}) ) \nonumber \\
&-\frac{\Delta \tau}{g_e^2} E_{ij}^\tau (\vec{r}) E_{ij}^\tau (\vec{r}) .  \nonumber \\
  +i (A_0^\tau(\vec{r})-A_0^\tau(\vec{r}+\mathbf{i})&-A_0^\tau(\vec{r}+\mathbf{j})\nonumber \\
  +A_0^\tau(\vec{r}+\mathbf{i}+\mathbf{j})  +A_{ij}^\tau &(\vec{r})-A_{ij}^{\tau+1} (\vec{r}) ) E_{ij}^\tau (\vec{r}) ) \nonumber
\end{align}
where we have made a redefinition of the measure $DA_{\tau}\equiv DA D\phi$.

Now using 
\be
\exp(z \cos(\phi)) \simeq \sum_n \exp \left( -\frac{1}{2z}n^2 + i \phi n \right).
\ee
to carry out the sum over electric field variables we find
\begin{align}
S = -\sum_{\tau,\vec{r},i<j}&\frac{2\Delta \tau}{g_m^2}\cos( A_{jk}(\vec{r}+\vec{i})-A_{jk}(\vec{r})\nonumber \\
&\qquad \qquad-A_{ki}(\vec{r}+\vec{j})+A_{ki}(\vec{r}))  \\
& +\frac{g_e^2}{2\Delta \tau}\cos (A_0^\tau(\vec{r})-A_0^\tau(\vec{r}+\vec{i})-A_0^\tau(\vec{r}+\vec{j})  \nonumber  \\
& \qquad \quad +A_0^\tau(\vec{r}+\vec{i}+\vec{j})+A_{ij}^\tau (\vec{r})-A_{ij}^{\tau+1} (\vec{r})) \nonumber
\end{align}
\par While this is the most general form of the action, we take the couplings constants for each therm to be identical $\frac{2\Delta \tau}{g_m^2}=\frac{g_e^2}{2\Delta \tau} \equiv \beta$ giving the gauge action in the main text. A similar calculation maybe carried out for the Higgs coupling leading to the final form of the lattice action.

\section{Pinch Point Correlations from the Continuum Field Theory}
\label{sec:weak_coupling_pinch_points}

In the main text, we discuss at length the pure gauge theory in the strong coupling limit as this is the limit most relevant to characterising the gauge theory in 4D for all $\beta$. It is interesting, though, to contrast these results with those obtained by taking a naive continuum limit where the lattice plaquette length scale $a\rightarrow 0$. 

Starting with lattice action:
\begin{align}
\mathcal{S} = -\beta\sum_{\tau}\sum_{\vec{r}}&\cos(\Delta_iA_{jk}(x) - \Delta_jA_{ki}(x))\nonumber \\
+& \cos(\Delta_i\Delta_jA_0(x) - \Delta_0A_{ij}(x)).
\end{align}
we obtain
\begin{align}
\mathcal{S} = \frac{\beta}{2}\sum_{\tau}\sum_{\vec{r}}&(\Delta_iA_{jk}(x) - \Delta_jA_{ki}(x))^2\nonumber \\
+& (\Delta_i\Delta_jA_0(x) - \Delta_0A_{ij}(x))^2
\end{align}
to second order in the gauge fields. 

We choose a gauge in which the ``divergence-free" condition is satisfied at each vertex,
\begin{align}
&-(A_0(x^{\mu}) - A_0(x^{\mu}-e^0))) + (A_{yz}(x^{\mu})  \nonumber\\
&- A_{yz}(x^{\mu}-e^{2})- A_{yz}(x^{\mu}-e^{3}) + A_{yz}(x^{\mu}-e^{2}-e^{3}))\nonumber\\
&+ (A_{zx}(x^{\mu}) - A_{zx}(x^{\mu}-e^{3})\nonumber \\
&- A_{zx}(x^{\mu}-e^{1}) + A_{x}(x^{\mu}-e^{3}-e^{1}))\nonumber\\
&+ (A_{xy}(x^{\mu}) - A_{xy}(x^{\mu}-e^{1})\nonumber\\
&- A_{xy}(x^{\mu}-e^{2}) + A_{xy}(x^{\mu}-e^{1}-e^{2})) ) = 0
\label{eq:Fracton_gaugecondition}
\end{align}
This is the direct generalization of Feynman gauge in the ordinary $U(1)$ case to this model.

Fourier transforming via
\begin{eqnarray}
\begin{cases}
A_{jk}(p_{\mu}) = \frac{1}{\sqrt{N^4}}\sum_{p_{\mu}}e^{-ip_{\mu}x^{\mu}}A_{jk}(x^{\mu})\\
A_{0}(p_{\mu}) = \frac{1}{\sqrt{N^4}}\sum_{p_{\mu}}e^{-ip_{\mu}x^{\mu}}A_{0}(x^{\mu})
\end{cases}.
\end{eqnarray}
the gauge constraint (\ref{eq:Fracton_gaugecondition}) becomes
\begin{align}
&-(1-e^{-ip_0})A_0(p_{\mu}) + (1-e^{-ip_2})(1-e^{-ip_3})A_{23}(p_{\mu})\nonumber \\
& + (1-e^{-ip_3})(1-e^{-ip_1})A_{31}(p_{\mu}) \nonumber \\
&+ (1-e^{-ip_1})(1-e^{-ip_2})A_{12}(p_{\mu}) = 0.
\label{eq:Fracton_gaugecondition_momentum}
\end{align}

The quadratic action is then
\begin{align}
\mathcal{S} =& \frac{\beta}{2}\begin{bmatrix} A_0(-p_{\mu}) & A_{23}(-p_{\mu}) & A_{31}(-p_{\mu}) & A_{12}(-p_{\mu}) \end{bmatrix} \nonumber \\
&D^{-1}(p_{\mu}) 
\begin{bmatrix} 
A_0(p_{\mu}) \\
A_{23}(p_{\mu}) \\ 
A_{31}(p_{\mu}) \\ 
A_{12}(p_{\mu}) \\
\end{bmatrix}
\end{align}
with propagator
\begin{align}
[D^{-1}(p_{\mu})]_{00} & = |e^{ip_1}-1|^2|e^{ip_2}-1|^2 + |e^{ip_2}-1|^2|e^{ip_3}-1|^2\nonumber \\
&+ |e^{ip_3}-1|^2|e^{ip_1}-1|^2 \nonumber \\
[D^{-1}(p_{\mu})]_{10} & = -(e^{ip_2}-1)(e^{ip_3}-1)(e^{-ip_0}-1) \nonumber \\
[D^{-1}(p_{\mu})]_{11} & = |e^{ip_0}-1|^2 + 2|e^{ip_1}-1|^2 \nonumber \\
[D^{-1}(p_{\mu})]_{20} & = -(e^{ip_3}-1)(e^{ip_1}-1)(e^{-ip_0}-1) \nonumber \\
[D^{-1}(p_{\mu})]_{21} & = -(e^{-ip_2}-1)(e^{ip_1}-1)  \nonumber \\
[D^{-1}(p_{\mu})]_{22} & = |e^{ip_0}-1|^2 + 2|e^{ip_2}-1|^2 \nonumber \\
[D^{-1}(p_{\mu})]_{30} & = -(e^{ip_1}-1)(e^{ip_2}-1)(e^{-ip_0}-1) \nonumber \\
[D^{-1}(p_{\mu})]_{31} & =  -(e^{-ip_3}-1)(e^{ip_1}-1) \nonumber \\
[D^{-1}(p_{\mu})]_{32} & = -(e^{-ip_3}-1)(e^{ip_2}-1) \nonumber \\
[D^{-1}(p_{\mu})]_{33} & =  |e^{ip_0}-1|^2 + 2|e^{ip_3}-1|^2
\end{align}
We define its eigenvalues $\varepsilon_m(p_{\mu})$ and eigenvectors $\vec{u}^{(m)}(p_{\mu})$ by the usual relation,
\begin{eqnarray}
D^{-1}(p_{\mu})\vec{u}^{(m)}(p_{\mu}) = \varepsilon_m(p_{\mu})\vec{u}^{(m)}(p_{\mu}).
\end{eqnarray}
Out of the 4 eigenmodes, $m=0, 1, 2, 3$, the lowest mode ($m=0$) gives the zero mode: $\varepsilon_0(p_{\mu})=0$, as it corresponds to the motion of the gauge fields in the gauge-equivalent space.
The corresponding eigenvector can be obtained from Eq.~(\ref{eq:Fracton_gaugecondition_momentum}) as
\begin{eqnarray}
\vec{u}^{(0)}(p_{\mu}) = \begin{bmatrix}
A_0(p_{\mu})\\
A_{23}(p_{\mu})\\
A_{31}(p_{\mu})\\
A_{12}(p_{\mu})
\end{bmatrix}
=
\begin{bmatrix}
-(1-e^{ip_0})\\
(1-e^{ip_2})(1-e^{ip_3})\\
(1-e^{ip_3})(1-e^{ip_1})\\
(1-e^{ip_1})(1-e^{ip_2})
\end{bmatrix}.
\end{eqnarray}
Expressions for the other eigenvectors cannot be found explicitly. We write the normal modes,
\begin{align}
\phi_m(p_{\mu})\equiv& u^{(m)*}_0(p_{\mu})A_0(p_{\mu}) + u^{(m)*}_{23}(p_{\mu})A_{23}(p_{\mu})\nonumber \\
&+ u^{(m)*}_{31}(p_{\mu})A_{31}(p_{\mu}) + u^{(m)*}_{12}(p_{\mu})A_{12}(p_{\mu}),
\end{align}
among which $\phi_0(p_{\mu})$ vanishes identically, due to the gauge condition, Eq.~(\ref{eq:Fracton_gaugecondition_momentum}).
Then the gauge fields can be expanded in the physical space as
\begin{eqnarray}
\begin{cases}
A_0(p_{\mu}) = \sum_{m=1}^3u^{(m)}_0(p_{\mu})\phi_m(p_{\mu})\\
A_{ij}(p_{\mu})= \sum_{m=1}^3u^{(m)}_{ij}(p_{\mu})\phi_m(p_{\mu})
\end{cases}.
\end{eqnarray}
With the normal modes, we can write down the action as
\begin{align}
&\mathcal{S} = \frac{\beta}{2}\sum_{p_{\mu}}\sum_{m=1}^3\varepsilon_m|\phi_m(p_{\mu})|^2 = \beta\sum_{p_{\mu}>0}\sum_{m=1}^3\varepsilon_m|\phi_m(p_{\mu})|^2.
\end{align}

We are now in a position to compute correlation functions of operators. We focus on the time space cube 
\begin{align}
 L_{ij\tau}(x^{\mu}) \equiv& U_{\tau}(x^{\mu})U^{-1}_{\tau}(x^{\mu}+e^i)U^{-1}_{\tau}(x^{\mu}+e^j)U_{\tau} \nonumber \\&(x^{\mu}+e^i+e^j)U_{\tau}(x^{\mu})U^{-1}_{\tau}(x^{\mu}+e^0)\\
=\exp\Bigl[i&\bigg\{(A_0(x^{\mu}) - A_0(x^{\mu}+e^i) - A_0(x^{\mu}+e^j)\nonumber \\
+ &A_0(x^{\mu}+e^i+e^j) + (A_{23}(x^{\mu}) - A_{23}(x^{\mu}+e^0))\bigg\}\Bigr] \nonumber
\end{align}
and, for concreteness, we take $(i,j)=(2,3)$ (the results can be generalized for other $(i,j)$ by symmetry arguments). In momentum space this is
\begin{align}
L_{23\tau}(x^{\mu}) 
&= \exp\Bigl[i\frac{1}{\sqrt{N^4}}\sum_{p_{\mu}}\sum_{m=1}^3e^{ip_{\mu}x^{\mu}}
\nonumber \\
\bigg\{
(1-e^{ip_2})&(1-e^{ip_3})u^{(m)}_0(p_{\mu})\phi_m(p_{\mu}) + (1-e^{ip_0})u^{(m)}_{23}(p_{\mu})\phi_{m}(p_{\mu})
\bigg\}\Bigr].
\end{align}
We find:
\begin{align}
&\langle L_{23\tau}(x^{\mu})\rangle =\nonumber \\
&\exp\Bigl[-\frac{1}{N^4}{\sum_{p_{\mu},m}\frac{|(1-e^{ip_2})(1-e^{ip_3})u^{(m)}_0(p_{\mu}) + (1-e^{ip_0})u^{(m)}_{23}(p_{\mu})|^2}{\beta\varepsilon_m(p_{\mu})}\Bigr]} \nonumber \\
&\equiv \exp (-B/\beta).
\end{align}
and
\begin{align}
  &\langle L^{\dagger}_{23\tau}(x^{\mu})L_{23\tau}(0)\rangle  = \nonumber \\
  &\exp\left[ -\frac{1}{N^4}\sum_{p_{\mu}}\sum_{m=1,2,3}\frac{1}{2\beta\varepsilon_m}|1-e^{ip_{\mu}x^{\mu}}|^2\nonumber\right. \\
  & \qquad \quad|(1-e^{ip_2})(1-e^{ip_3})u^{(m)}_0(p_{\mu})  \nonumber  \left.  + (1-e^{ip_0})u^{(m)}_{23}(p_{\mu})|^2\right].
\end{align}
It follows that the time tube expectation value goes like
\begin{align}
\langle L_{23}(x^{\mu})\rangle = \exp(-N_{\tau}B/\beta).
\end{align}

and the extended time tube goes like

\begin{equation}
\left\langle L_{23}(x^{\mu};r) \right\rangle =\exp \left(  - \frac{r N_t B}{\beta}  \right)
 \end{equation}

We now consider the connected correlator 
\begin{eqnarray}
C_{23\tau}(x^{\mu})\equiv\langle L^{\dag}_{23\tau}(x^{\mu})L_{23\tau}(0)\rangle - \langle L_{23\tau}(0)\rangle^2.
\end{eqnarray}
which, in momentum space, is 
 \begin{align}
C_{23\tau}(p^{\mu}) = \frac{1}{\beta}\sum_{m=1,2,3}&\frac{1}{\varepsilon_m}|(1-e^{ip_2})(1-e^{ip_3})u^{(m)}_0(p_{\mu})\nonumber \\
&+ (1-e^{ip_0})u^{(m)}_{23}(p_{\mu})|^2
\end{align}
We plot $C_{23\tau}(p^{\mu})$ for different momentum slices.

\begin{figure}[h]
\centering
    \includegraphics[width=0.8\columnwidth]{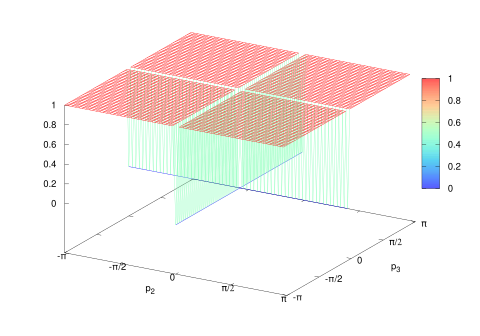}
    \caption{Four-fold pinch points for $p_0=p_1=0$ and plotted in the $p_2,p_3$ plane. }	
    \label{fig:pinchpoint1}
\end{figure}

\begin{figure}[h]
\centering
    \includegraphics[width=0.8\columnwidth]{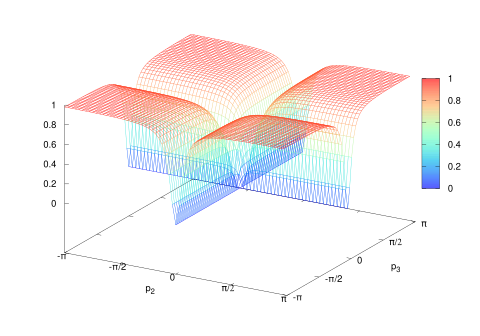}
    \caption{Plane parallel to the four-fold pinch points for $p_0=0$, $p_1=\pi/16$ and plotted in the $p_2,p_3$ plane showing how the pinch points are broadened into valleys.}	
    \label{fig:pinchpoint2}
\end{figure}

Fig.~\ref{fig:pinchpoint1} shows the $p_2, p_3$ plane for fixed $p_0=p_1=0$. In this plane there is a four-fold symmetry with a singular line along the $p_2=0$ and $p_3=0$ lines. Shifting the plane by plotting instead at fixed $p_1=\pi/16$ reveals that where there were singular features there are valleys along these same lines (Fig.~\ref{fig:pinchpoint2}). The pinch points are localized in this plane. 

\section{Symmetries of Hollow Rank-2 U(1)}
 \label{sec:symmetries}

 \par We now verify the symmetries associated with the hollow rank-2 U(1) action

\begin{align}
	S_{\rm Gauge} = -\beta \sum_{\tau=1}^N  \sum_{\vec{r}}& \cos( \Delta_i A_{jk}(\vec{r})-\Delta_j A_{ki}(\vec{r}) ) \nonumber \\
    &+ \cos (\Delta_i \Delta_j A_0^\tau(\vec{r}) -\Delta_0 A_{ij}^\tau (\vec{r})).
\end{align}

\par First let us verify how it behaves with rotations. The figure below sketches how do the plaquette elements of the action vary with $90^{\circ}$ rotations along some axis. The ``orientation" of each plaquette is represented with an arrow.

\begin{figure}[h]
\centering
    \includegraphics[width=0.5\columnwidth]{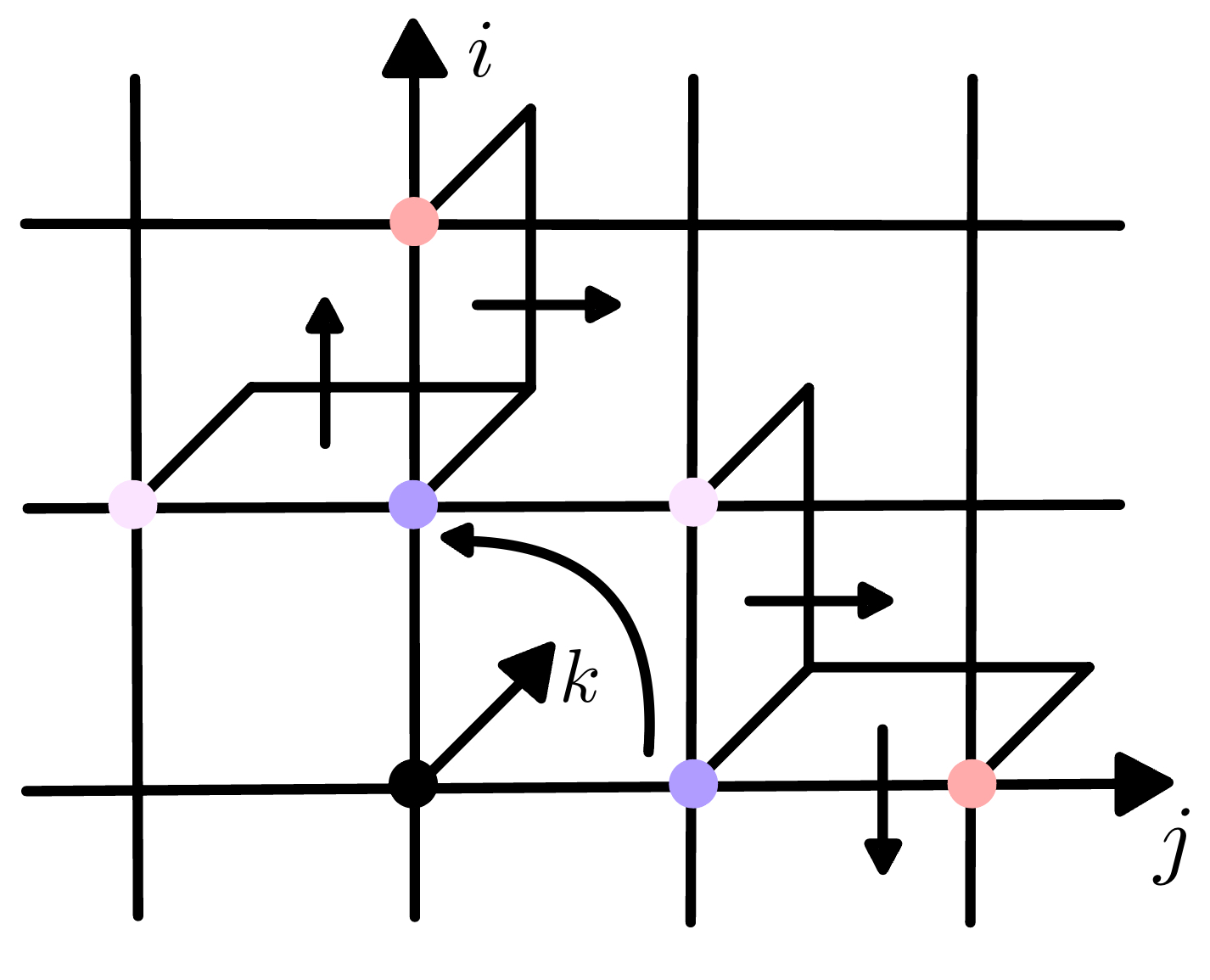}
    \caption{Changes due to $90^{\circ}$ rotation around the $k$ axis}	
\end{figure}

\par From the figure it is clear that rotations around the $k$ axis affect plaquettes in the following manner

\begin{equation}
	A_{ij} \rightarrow A_{ij} \qquad A_{ik} \rightarrow -A_{jk} \qquad A_{jk} \rightarrow A_{ik},
\end{equation}
where we need to take into the account the cyclic nature of the indices when replacing them by actual coordinates $ijk=\{xyz,yzx,zxy\}$.
\par The relative positions between the colored dots also make it clear that the spatial differences need to transform like 

\begin{equation}
	\Delta_{k} \rightarrow \Delta_{k} \qquad \Delta_{i} \rightarrow -\Delta_{j} \qquad \Delta_{j} \rightarrow \Delta_{i}.
\end{equation}

\par The temporal links live in sites in the corners of the plaquettes and have alternating signs as we go around a plaquette. Upon rotations their positions in the plaquettes will also rotate. This implies that upon rotations the time link variables should change sign. The differences in the temporal direction are unaffected.

\begin{equation}
	A_0 \rightarrow -A_0 \qquad \Delta_0 \rightarrow \Delta_0.
\end{equation}

\par By taking all this into consideration it is straightforward to see that the action is invariant under $90^{\circ}$ rotations around any axis. 

\par Now we can look into reflection symmetries. While these are easier to visualize, we present a visual representation of reflections on the gauge variables for clarity.

\begin{figure}[h]
\centering
    \includegraphics[width=0.6\columnwidth]{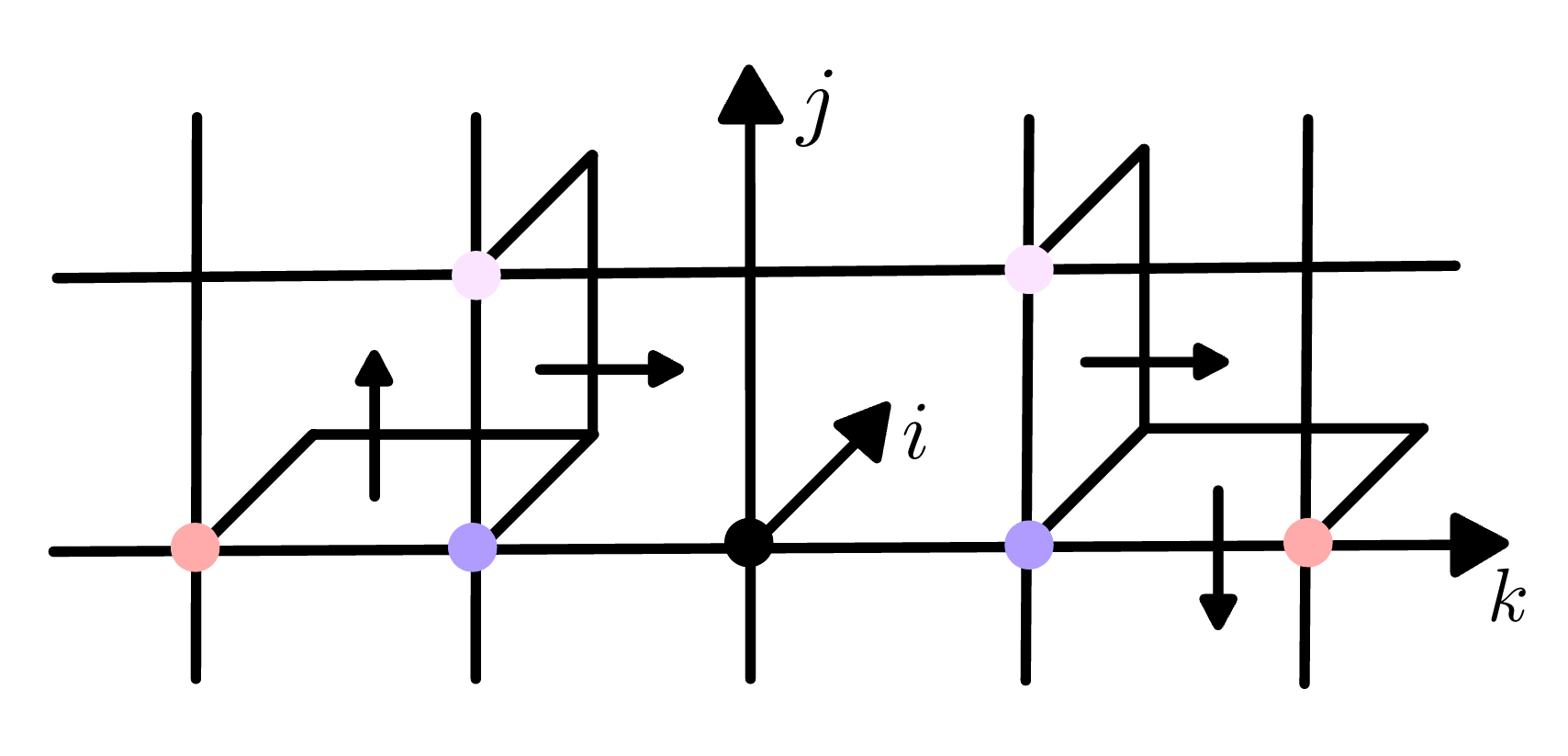}
    \caption{Changes due to a reflection from $k$ to $-k$}	
\end{figure}

\par The effects of reflection from $k$ to $-k$ on the plaquettes are then

\begin{equation}
	A_{ij} \rightarrow A_{ij} \qquad A_{ik} \rightarrow -A_{ik} \qquad A_{jk} \rightarrow -A_{jk}
\end{equation}

and on the differences 

\begin{equation}
	\Delta_{k} \rightarrow -\Delta_{k} \qquad \Delta_{i} \rightarrow \Delta_{i} \qquad \Delta_{j} \rightarrow \Delta_{j}.
\end{equation}

\par For this type of transformation both the time links and the temporal differences remain unchanged 

\begin{equation}
	A_0 \rightarrow A_0 \qquad \Delta_0 \rightarrow \Delta_0.
\end{equation}

\par Again, it is simple to see that the action is invariant for reflections around any direction.

\subsection{Correlation Functions}

\par While the action is invariant to reflections and translation, the same is not necessarily true for the correlation functions $\langle L_{ij\tau}( \vec{r_1})L_{ij\tau}^{\dag}( \vec{r_2}) \rangle $ and $\langle L_{ik\tau}( \vec{r_1})L_{kj\tau}^{\dag}( \vec{r_2}) \rangle $.

\par From the earlier results we can deduce that the operators $L_{ij\tau}$ transform in the following way for rotations 

\begin{equation}
	L_{ij} \xrightarrow{k^{90^{\circ}}} L_{ij} \qquad L_{ij} \xrightarrow{i^{90^{\circ}}} L_{ki}^{\dag}  \qquad  L_{ij} \xrightarrow{j^{90^{\circ}}} L_{jk}.
\end{equation}

\par This then implies that for the correlation operators we have 

\begin{align}
	&L_{ij} L_{ij}^{\dag} \xrightarrow{k^{90^{\circ}}} L_{ij} L_{ij}^{\dag} \\
	 L_{ij} L_{ij}^{\dag}& \xrightarrow{i^{90^{\circ}}} L_{ki}^{\dag} L_{ki} \xrightarrow{i^{90^{\circ}}}L_{ij}^{\dag}L_{ij} \nonumber \\
       L_{ij} L_{ij}^{\dag}& \xrightarrow{j^{90^{\circ}}} L_{jk}L_{jk}^{\dag}\xrightarrow{j^{90^{\circ}}} L_{ij}^{\dag}L_{ij}
\end{align}

\par and 

\begin{align}
	L_{ik} L_{kj}^{\dag} \xrightarrow{k^{90^{\circ}}} L_{kj}^{\dag} L_{ki}^{\dag}& \xrightarrow{k^{90^{\circ}}} L_{ik}^{\dag} L_{kj}\\
	L_{ik} L_{kj}^{\dag} \xrightarrow{i^{90^{\circ}}} L_{ij}  L_{kj}^{\dag} \xrightarrow{i^{90^{\circ}}}L_{ik}^{\dag} L_{kj}^{\dag}&\xrightarrow{i^{90^{\circ}}} L_{ij}^{\dag}  L_{kj}^{\dag}\xrightarrow{i^{90^{\circ}}}L_{ik}L_{kj}^{\dag} \\
	   L_{ik} L_{kj}^{\dag} \xrightarrow{j^{90^{\circ}}} L_{ik}L_{ij}\xrightarrow{j^{90^{\circ}}} L_{ik}L_{kj}&\xrightarrow{j^{90^{\circ}}} L_{ik}L_{ij}^{\dag}\xrightarrow{j^{90^{\circ}}} L_{ik}L_{kj}^{\dag}.
\end{align}

\par The symmetries present in the theory do not carry on to the correlation functions. For $\langle L_{ij\tau}( \vec{r_1})L_{ij\tau}^{\dag}( \vec{r_2}) \rangle $ we keep the four-fold symmetry around the $k$ axis, but it breaks down to two-fold symmetry for the $i$ and $j$ axis. The lack of symmetry is even more prevalent for $\langle L_{ik\tau}( \vec{r_1})L_{kj\tau}^{\dag}( \vec{r_2}) \rangle $, where there is only two-fold symmetry around $k$.

\par A similar analysis can be done for reflections. The elementary operators $L_{ij\tau}$ transform like

\begin{equation}
	L_{ij} \xrightarrow{k\rightarrow-k} L_{ij} \qquad L_{ij} \xrightarrow{i\rightarrow-i} L_{ij}^{\dag}  \qquad  L_{ij} \xrightarrow{j\rightarrow-j} L_{ij}^{\dag}.
\end{equation}

\par This then implies that for the correlation operators we have 

\begin{align}
	L_{ij}L_{ij}^{\dag} \xrightarrow{k\rightarrow-k}& L_{ij}L_{ij}^{\dag} \qquad L_{ij}L_{ij}^{\dag} \xrightarrow{i\rightarrow-i} L_{ij}^{\dag}L_{ij}\nonumber \\
    & L_{ij}L_{ij}^{\dag} \xrightarrow{j\rightarrow-j} L_{ij}^{\dag}L_{ij}
\end{align}

\par and 

\begin{align}
	L_{ik}L_{kj}^{\dag} \xrightarrow{k\rightarrow-k} &L_{ik}^{\dag}L_{kj} \qquad L_{ik}L_{kj}^{\dag} \xrightarrow{i\rightarrow-i} L_{ik}^{\dag}L_{kj}^{\dag} \nonumber \\
    &L_{ik}L_{kj}^{\dag} \xrightarrow{j\rightarrow-j} L_{ik}L_{kj}.
\end{align}

\par For reflections the symmetries prevail for $\langle L_{ij\tau}( \vec{r_1})L_{ij\tau}^{\dag}( \vec{r_2}) \rangle $ but, again, the operator $\langle L_{ik\tau}( \vec{r_1})L_{kj\tau}^{\dag}( \vec{r_2}) \rangle$ proves to have a higher degree of assymetry since the reflection symmetries around $i$ and $j$ break down.

\par A summary of these findings can be found in the table below.

\begin{table}[h]
\centering
\begin{tabular}{|l|l|l|}
\hline
Operator & Rotation & Reflection \\ \hline
    $L_{ij}L_{ij}^{\dag}$     &    $k^{90^{\circ}}$,$i^{180^{\circ}}$ and $j^{180^{\circ}}$      &      All axis      \\ \hline
   $L_{ik}L_{kj}^{\dag}$      &    $k^{180^{\circ}}$      &      $k\rightarrow-k$      \\ \hline
\end{tabular}
\end{table}

\bibliography{references}

\end{document}